\documentclass[fleqn,usenatbib,useAMS]{mnras}
\usepackage{amsmath}	
\usepackage{txfonts}
\usepackage[T1]{fontenc}
\usepackage{ae,aecompl}
\usepackage{graphicx,color}	
\usepackage{amssymb}	
\usepackage{amsbsy}
\usepackage[normalem]{ulem}
\usepackage{mathrsfs,bm,xspace}
\newcommand{\bcdot}{\ensuremath{%
  \mathchoice%
   {\mskip\thinmuskip\lower0.2ex\hbox{\scalebox{1.5}{$\cdot$}}\mskip\thinmuskip}}%
   {\mskip\thinmuskip\lower0.2ex\hbox{\scalebox{1.5}{$\cdot$}}\mskip\thinmuskip}%
   {\lower0.3ex\hbox{\scalebox{1.2}{$\cdot$}}}%
   {\lower0.3ex\hbox{\scalebox{1.2}{$\cdot$}}}%
}

\usepackage[normalem]{ulem}
\usepackage[usenames]{xcolor}

\newcommand{\Mstream}{{\it M-streaming}\xspace}
\newcommand{\Mflatturb}{{\it M-turbulence}\xspace}
\newcommand{\Mprimary}{{\it M-primaries}\xspace}

\newcommand{\dd}{\mathrm{d}}
\newcommand{\Vph}{\varv_\mathrm{ph}}

\newcommand{\RH}{R_\rmn{RH}}
\newcommand{\bvel}{\ensuremath{\boldsymbol{\varv}}}
\newcommand{\bnabla}{\ensuremath{\boldsymbol{\nabla}}}
\newcommand{\eb}{\epsilon_B}

\newcommand{\p}{\rmn{p}}
\newcommand{\kB}{k_{\rmn{B}}}
\newcommand{\eps}{\varepsilon}
\newcommand{\CR}{\rmn{CR}}

\definecolor{mygreen}{rgb}{0.,0.5,0.}

\title[Origin of Seed Electrons]{Turbulence and Particle Acceleration in Giant Radio Haloes: the Origin of Seed Electrons}  

\author[A. Pinzke, S. Peng Oh and C. Pfrommer] 
{Anders Pinzke$^{1,2}$\thanks{apinzke@fysik.su.se (AP); peng@physics.ucsb.edu (SPO); christoph.pfrommer@h-its.org (CP)}, S. Peng Oh$^{3}$\footnotemark[1], and Christoph Pfrommer$^{4}$\footnotemark[1]\\
$^{1}$The Oskar Klein Centre for Cosmoparticle Physics, Stockholm University, AlbaNova University Center, SE - 106 91
  Stockholm, Sweden\\
$^{2}$Dark Cosmology Center, University of Copenhagen,
  Juliane Maries Vej 30, DK-2100 Copenhagen, Denmark\\
  $^{3}$University of California - Santa Barbara,
  Department of Physics, CA 93106-9530, USA\\
$^{4}$Heidelberg Institute for Theoretical Studies
  (HITS), Schloss-Wolfsbrunnenweg 35, 69118 Heidelberg, Germany}

\begin{document}
\pagerange{\pageref{firstpage}--\pageref{lastpage}} \pubyear{2015}
\maketitle
\label{firstpage}



\begin{abstract}
  About $1/3$ of X-ray-luminous clusters show smooth, Mpc-scale radio
  emission, known as giant radio haloes. One promising model for radio
  haloes is Fermi-II acceleration of seed relativistic electrons by
  compressible turbulence. The origin of these seed electrons has
  never been fully explored. Here, we integrate the Fokker-Planck
  equation of the cosmic ray (CR) electron and proton distributions
  when post-processing cosmological simulations of cluster formation,
  and confront them with radio surface brightness and spectral data of
  Coma. For standard assumptions, structure formation shocks lead to a
  seed electron population which produces too centrally concentrated
  radio emission. Matching observations requires modifying properties
  of the CR population (rapid streaming; enhanced CR electron
  acceleration at shocks) or turbulence (increasing
  turbulent-to-thermal energy density with radius), but at the expense
  of fine-tuning. In a parameter study, we find that radio properties
  are exponentially sensitive to the amplitude of turbulence, which is
  inconsistent with small scatter in scaling relations. This
  sensitivity is removed if we relate the acceleration time to the
  turbulent dissipation time. In this case, turbulence above a
  threshold value provides a fixed amount of amplification;
  observations could thus potentially constrain the unknown CR seed
  population. To obtain sufficient acceleration, the turbulent
  magneto-hydrodynamics cascade has to terminate by transit time
  damping on CRs, i.e., thermal particles must be scattered by plasma
  instabilities. Understanding the small scatter in radio halo scaling
  relations may provide a rich source of insight on plasma processes
  in clusters.
\end{abstract}


\section{Introduction}
About one third of X-ray-luminous clusters show smooth, unpolarised radio
emission on $\sim$Mpc scales, known as giant radio haloes (RHs)
\citep{2014IJMPD..2330007B}. They appear only in disturbed, merging clusters and
the RH luminosity correlates with the X-ray luminosity
\citep{2001A&A...369..441G,2012A&ARv..20...54F} and the Compton $y$-parameter
\citep{2012MNRAS.421L.112B,2013A&A...554A.140P}. The RHs show that CR electrons
and magnetic fields permeate a large volume fraction of the intra-cluster medium
(ICM). The dominant CR source, given the smoothness and enormous extent of RHs,
is thought to be structure formation shocks \citep{miniati01,pfrommer08}. At the
same time, plasma processes, the origin of magnetic fields and particle
acceleration in a turbulent, high-$\beta$ plasma (in which the thermal pressure
predominates the magnetic pressure) like the ICM are not well understood. Radio
haloes thus provide an incisive probe of non-thermal processes in the ICM.

There have been two competing models proposed to explain RHs.  The
radio emitting electrons in the ``hadronic model'' are produced in
inelastic (hadronic) CR proton interactions with protons of the
ambient thermal ICM, which generates pions that eventually decay into
electrons and positrons, depending of the charge of the initial pion
(\citealp{1980ApJ...239L..93D,1999APh....12..169B,2001ApJ...562..233M,
  2004A&A...413...17P,2008MNRAS.385.1211P,ensslin11}). CR protons and
heavier nuclei may have been accelerated and injected into the ICM by
structure formation shocks, active galactic nuclei and galactic
winds. However, the strong bimodality that separates X-ray luminous
clusters into radio-active and radio-quite clusters (requiring a
fast switch on/off mechanism of the RH emission) and the very extended
RH emission at low frequencies in Coma (352~MHz) represent a major
challenge to this model class \citep{brunetti12,2014MNRAS.438..124Z}.

The alternative model for RHs is re-energisation of seed suprathermal
electrons by Fermi II acceleration when ICM turbulence becomes
transsonic during mergers
\citep{1987A&A...182...21S,1993ApJ...406..399G,2001MNRAS.320..365B,
  2004MNRAS.350.1174B,brunetti07,brunetti11,miniati15}. Due
to the short radiative cooling time of high-energy relativistic
electrons, the cluster synchrotron emission quickly fades away after a
merger, which naturally explains the observed bimodality of RHs
\cite[see e.g.][]{2013MNRAS.429.3564D,2014MNRAS.443.3564D}.

However, there is a salient piece missing in the turbulent
reacceleration model. It relies heavily on the assumption of an
abundant, volume-filling population of seed suprathermal electrons;
direct Fermi II acceleration from the thermal pool is precluded by
strong Coulomb losses
\citep{2008ApJ...682..175P,2012ApJ...759..113C}. These seeds are
presumed to be either fossil CR electrons (CRes) accelerated by
diffusive shock acceleration (DSA) during structure formation
\citep{1999ApJ...520..529S}, or secondaries injected by hadronic
interaction of CR protons (CRps) with thermal protons
\citep{brunetti11}.

\begin{figure}
  \includegraphics[width=1.0\columnwidth]{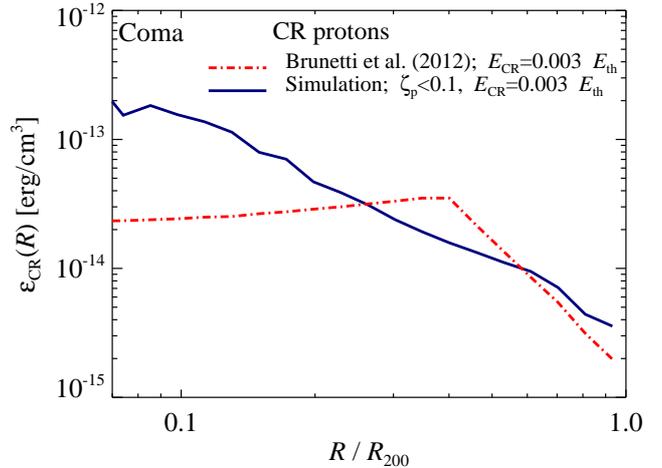}
  \caption{Spatial distribution of CRp energy density in the Coma
    cluster. The red dash-dotted line shows the required distribution
    of seed CRps that generate secondary electrons via proton-proton
    (p-p) collisions required to reproduce Coma radio brightness
    observations after Fermi-II reacceleration \citep{brunetti12}. The
    blue solid line shows the distribution of fossil CRps found in
    cosmological simulations, which disagrees with the required
    profile. To better compare the two models in this figure, we
    normalise the required distribution of CRps by fixing the total
    CRp energy $E_\rmn{CR}$ to 0.3 per cent of the total thermal
    energy, consistent with observations
    \citep{2014ApJ...787...18A,2012ApJ...757..123A}.}
  \label{fig:Edens}
\end{figure}

While analytic estimates have been made, there has been no ab initio
demonstration that structure formation can lead to the required
abundance of seed electrons with the correct spatial and spectral
characteristics. This is a non-trivial requirement: Coulomb cooling in
dense cluster cores is severe, and DSA fossil electrons may not
survive. On the other hand, for secondaries to constitute the seed
population, the CRp population required in the best-studied case of
the Coma cluster must have a very broad and flat (or even slightly
inverted) spatial profile \citep{brunetti12}, in contrast with the
thermal plasma whose energy density declines steeply with radius. In
Figure~\ref{fig:Edens} we show that such a distribution is not predicted
by cosmological simulations \cite[see
  also][]{pinzke10,2014MNRAS.439.2662V}. If CRps are predominantly
advected with the cluster plasma, their distribution will be peaked
towards the cluster centre and show a similar characteristics as the
thermal plasma. As a consequence, the distribution of secondary
electrons and the resulting radio synchrotron emission is also peaked
since the hadronic reaction is a two-body scattering process. Hence,
the simulated emission falls short of the observed extended and flat
radio profile of the Coma cluster.

Indeed, arriving at a seed population with the required characteristics is
highly constraining, and has the potential to teach us much about the origin of
CRps/CRes in clusters.  In this work, we use our hydrodynamical zoom simulations
of galaxy clusters in a cosmological setting to follow the distribution
functions of seed populations for CRps and CRes, and integrate the Fokker-Planck
equation of CR transport along Lagrangian particle trajectories. We model
diffusive shock acceleration at structure formation shocks, and account for
various loss processes of CRs. Utilising new insights from our recent work on
DSA generated fossil electrons \citep{pinzke13}, we generate the first
quantitative calculation of primary and secondary seed electrons.

To compare this to observations, we model second-order Fermi acceleration by CR
interactions with magnetised turbulence. However, we assume a simplified and
stationary model for magnetic fields and turbulence. We do not account for the
time-varying energy density in compressible waves, which are thought to be
necessary for the acceleration process \citep{brunetti07,brunetti11}, as the
cluster merger proceeds.  So our approach is orthogonal (and complementary) to
e.g., simulations of \citealp{miniati15} that focus on the time-dependent
compressible turbulence while adopting a simplified treatment of CR. Our
approach of parametrizing turbulence enables us to vary parameters associated
with the spatial profile and the overall amplitude of compressible waves (that
can in principle vary depending on the details of a particular cluster merger).

In this paper, we explore how the radio surface brightness profile and spectrum
of the best known radio halo, Coma, can be used to constrain the underlying
properties of the seed CRs and turbulence. We aim to constrain the normalisation
and spatial profile of these two input ingredients in turbulent reacceleration
models. The outline of this paper is as follows. In Section~\ref{sect:method}, we
outline the basic physics of turbulent reacceleration of CRs which we use. In
Section~\ref{sec:results}, we use cosmological simulations to generate a seed CR
population, and combine it with our parametrized model of turbulence to produce
radio surface brightness profiles and spectra of Coma. We find that it is
possible to fit the observations using physically motivated modifications of the
seed population (rapid streaming; enhanced CR electron acceleration at shocks)
or turbulence (increasing the turbulent-to-thermal energy density, $\eps_{\rm
  turb}/\eps_{\rm th}$ with radius), but only at the expense of fine
tuning. In Section~\ref{sect:param_comp}, we explore the reason for this fine-tuning,
and seek ways to overcome it. We perform a parameter study on spherically
symmetric, static models where we vary properties of the seed population and
turbulence. We find exponential sensitivity to the amount of turbulence, which
can be eliminated if the turbulent acceleration and dissipation time are
linked. In this case, turbulence above a threshold value required to overcome
cooling provides a fixed amount of amplification; observations could then
potentially constrain the unknown CR seed population. We summarise and conclude
in Section~\ref{sec:conclusions}.

\section{Cosmic ray transport}
\label{sect:method} 

The transport of relativistic electrons and protons across cosmic time
into galaxy clusters is a complex problem that depends on the velocity
field of the gas (and its thermodynamic properties such as density,
temperature, and pressure) as well as non-thermal processes
(turbulence, magnetic fields, fossil CRs). We use high resolution
galaxy cluster simulations to derive the thermal and fossil CR
properties \citep[shock accelerated primary CRes and CRps, as well as
  secondary CRes produced in p-p collisions,
  see][]{2007MNRAS.378..385P,pfrommer08,pinzke10,pinzke13}.

\subsection{Basic equations}
As previously noted, secondaries produced by shock accelerated CRp
have the wrong spatial profile to explain RH observations. Because
they arise from a two body process, they are too centrally
concentrated. They also produce gamma-ray emission in excess of
Fermi-LAT upper limits
\citep{2012ApJ...757..123A,brunetti12,2014ApJ...787...18A,Ahnen2016}. 

Given a seed population of CRs, we adopt essentially the same set of
plasma physics assumptions as the reacceleration model for RHs
\citep{brunetti07,brunetti11}. We solve the isotropic, gyro-phase
averaged Fokker-Planck equation (via a Crank-Nicholson scheme) for the
time evolution of the CRe distribution in the Lagrangian frame
\citep{brunetti07,brunetti11}:
\begin{eqnarray}
{{d f_{\rmn{e}}(p,t)}\over{d t}} &\!=&
\frac{\partial}{\partial p}
\left\{
f_{\rmn{e}}(p,t)\left[
\left|\frac{d p}{d t}\right|_{\rm Coul} 
+ \frac{p}{3}\left(\bnabla\bcdot \bvel\right)
+ \left|\frac{d p}{d t}\right|_{\rm rad}\right.\right.
\nonumber\\
&-& \left.\left.{1\over{p^2}}{{\partial }\over{\partial p}}\left(p^2 D_{pp}\right) 
\right]\right\} - \left(\bnabla\bcdot \bvel\right) f_{\rmn{e}}(p,t)
\nonumber\\
&+& {{\partial^2 }\over{\partial p^2}}
\left[
D_{pp} f_{\rmn{e}}(p,t) \right]+ Q_{\rmn{e}}\left[p,t;f_{\rmn{p}}(p,t)\right]   \, .
\label{elettroni}
\end{eqnarray}
Here $f_{\rmn{e}}$ is the one-dimensional distribution in position $x$
(suppressed for clarity), momentum $p=P/(m_{\rmn{e}} c)$ and time $t$
(which is normalised such that the number density is given by
$n_{\rmn{CRe}}(t)=\int d p f_{\rmn{e}}(p,t)$), $d/d t=\partial/\partial
t+\bvel\bcdot\bnabla$ is the Lagrangian derivative, $\bvel$ is the gas velocity,
$|d p/d t|$ represents Coulomb \citep[Coul,][]{1972Phy....60..145G} and
radiative \citep[rad,][]{1979rpa..book.....R} losses, respectively,
\begin{eqnarray}
   \left|\frac{d p}{d t}\right|_{\rm Coul} &=& \frac{3\,\sigma_\rmn{T}\,n_{\rmn{e}}\,c}{2\, \beta_{\rmn{e}}^2}
  \left[\ln\left(\frac{m_{\rmn{e}} c^2 \beta_{\rmn{e}} \sqrt{\gamma-1}}{\hbar\,\omega_\rmn{plasma}}\right)\right.\nonumber \\
    &-&\ln(2)\left(\frac{\beta_{\rmn{e}}^2}{2}+\frac{1}{\gamma}\right)+\frac{1}{2}+
    \left.\left(\frac{\gamma-1}{4\gamma}\right)^2\right]\,, \\
  \left|\frac{d p}{d t}\right|_{\rm rad~} &=& \frac{4}{3}\frac{\sigma_\rmn{T}}{m_{\rmn{e}}c}\,
  \frac{p^2}{\beta_{\rmn{e}}}\,\left[1+\left(\frac{B}{B_\rmn{CMB}}\right)^2\right]\,\eps_{\rmn{CMB}}\,.
\end{eqnarray}
Here $\beta_{\rmn{e}} = p/\sqrt{1+p^2}$ is the dimensionless velocity of CRs,
$\gamma=\sqrt{1+p^2}$ is the Lorentz factor of CRs,
$\omega_\rmn{plasma} = \sqrt{4\upi e^2 n_{\rmn{e}} / m_{\rmn{e}}}$ is the plasma
frequency, $n_e$ is the number density of free electrons, and
$\sigma_\rmn{T}= 8\upi e^4/3(m_{\rmn{e}} c^2)^2$ is the Thomson cross
section. The {\it rms} magnetic field strength is denoted by $B$ and
the equivalent field strength of the cosmic-microwave background is
given by $B_\rmn{CMB} = 3.24 (1 + z)^2\umu\rmn{G}$, where $z$ denotes
the redshift. In the peripheral cluster regions, where $B \ll
B_\rmn{CMB}$, the CRes loose virtually all their energy by means of
inverse Compton emission. $D_{pp}$ is the momentum space diffusion
coefficient (see Section~\ref{sec:reacc}), and
$Q_{\rmn{e}}$ denotes the injection rate of primary and secondary
electrons in the ICM (see section~\ref{sec:cosmo_sim}). The first term
containing the expression $\bnabla\bcdot \bvel$ represents Fermi-I
acceleration and the second term of this form describes adiabatic
gains and losses.

During post-processing of our Coma-like cluster simulation, we solve the
Fokker-Planck equation over a redshift interval from $z=5$ to 0. The simulated
cluster undergoes a major merger over the last 1-2~Gyrs that is thought to
inject large turbulent eddies. As is commonly assumed
\citep{brunetti07,brunetti11,2004ApJ...614..757Y,2013ApJ...771..131B} we assume
that about one Gyr after core passage the fields have decayed down to the
smallest scales $k_{\rm cut}$, and the radio halo turns on shortly after. We
choose this simulation snapshot to analyse. Note that in recent simulations by \citet{miniati15}, the turbulent reacceleration is strongest around core passage. However, we are not very sensitive to the adopted decay time, since the thermal and CR quantities are very similar a few 100 Myrs before and after $z = 0$, where we have chosen to evaluate the simulations.
In all our calculations we assume that turbulent reacceleration efficiently
accelerates particles for $\tau_{\rm cl} \sim 650$ Myrs (which is roughly the
cascade time on which turbulence is damped) and that during this turbulent phase
CR streaming and spatial diffusion can be neglected. In
Section~\ref{sect:self-limiting}, we explore sensitivity to the last assumption.

Thus far, we have ignored CR transport. However, if CRps stream in the ICM, then
their spatial profile could potentially flatten sufficiently
\citep{ensslin11,wiener13}. This scenario is very attractive: it generates seed
electrons with the right spatial footprint, and by removing CRps from the core,
obeys gamma-ray constraints. Turbulence plays two opposing roles: Alfv{\'e}nic
turbulence damps waves generated by the CR streaming instability
\citep{yan02,farmer04}, thus reducing self-confinement; but compressible fast
modes scatter CRs directly. Turbulent damping is still efficient for highly
subsonic conditions \citep{wiener13}, while we assume compressible fast modes to
only provide effective spatial confinement during the periods of transsonic,
highly super-Alfv{\'e}nic ($\mathcal{M}_{\rm A} \sim 5$) turbulence associated with
mergers. Thus, CRs can stream out when the cluster is kinematically
quiescent. Furthermore, even Alfv{\'e}nic streaming timescales are relatively
short \cite[$\sim 0.1-0.5$ Gyr;][]{wiener13} compared to the timescale on which
the CRp population is built up. Based on these findings, we adopt a toy model
for our \Mstream scenario in which CR streaming instantaneously produces flat
CRp profiles. We assume that CRs cannot stream significantly past perpendicular
$B$-fields at the accretion shock, so that the total number of CRs is conserved
within the virial radius during the streaming process. 

The time evolution of the spectral energy distribution of CRps,
$f_{\rmn{p}}$, is similarly given by:
\begin{eqnarray}
\lefteqn{
  {{d f_{\rmn{p}}(p,t)}\over{d t}} =
  {{\partial }\over{\partial p}}
  \left\{
  f_{\rmn{p}}(p,t)\left[ \left|{{d p}\over{d t}}\right|_{\rm Coul}
    + \frac{p}{3}\bnabla\bcdot \left(\bvel+\bvel_{\rmn{st}}\right)\right.\right.}
\nonumber\\
&&-\left.\left. {1\over{p^2}}{{\partial }\over{\partial p}}\left(p^2 D_{pp}\right)
\right]\right\} - f_{\rmn{p}}(p,t) \bnabla\bcdot \left(\bvel+\bvel_{\rmn{st}}\right) 
- \bvel_{\rmn{st}}\bcdot\bnabla f_{\rmn{p}}(p,t)
\nonumber\\
&&+\,\,\, {{\partial^2 }\over{\partial p^2}}
\left[ D_{pp} f_{\rmn{p}}(p,t) \right] - {{f_{\rmn{p}}(p,t)}\over{\tau_{\rm had}(p)}}
+ Q_{\rmn{p}}(p,t)\, ,
\label{eq:FP_p}
\end{eqnarray}
where $\bvel_{\rmn{st}}=-\varv_{\rmn{A}}\bnabla f_{\rmn{p}}/|\bnabla
f_{\rmn{p}}|$ is the streaming velocity in the isotropic transport
approximation, $\varv_{\rmn{A}}$ is the Alfv\'en speed,
and $Q_{\rmn{p}}(p,t)$ denotes the injection rate of shock accelerated
CRps as a function of momentum $p=P/(m_{\rmn{p}} c)$ and time $t$ (see
section~\ref{sec:cosmo_sim}). The timescale of
hadronic losses that produce pions via CRp collisions with thermal
protons of the ICM is given by 
\begin{eqnarray}
  \tau_{\rm had} = \left[c\,n_\rmn{th}\,\sigma^{+/-,0}(p)\right]^{-1}\,,
\end{eqnarray}
where we use the cross-section, $\sigma^{+/-,0}(p)$, given by the
fitting formula in \citet{1986ApJ...307...47D}.

\subsection{Turbulent reacceleration}
\label{sec:reacc}

In turbulent reacceleration, particles gain energy via Fermi II acceleration. Wave-particle energy exchange takes place via transit time damping \citep[TTD,][]{brunetti07,brunetti11}. The
TTD resonance requires the wave frequency to obey the resonance
condition, $\omega=k_\parallel\varv_\parallel$, where $k_\parallel$
and $\varv_\parallel$ are the parallel (projected along the magnetic
field) wavenumber of the compressible mode and particle velocity, respectively. This implies
that the particle transit time across the confining wave region
matches the wave period,
$\lambda_{\parallel}/\varv_{\parallel}=\tau_{\rmn{wave}}$ (hence the moniker, 'transit time damping'). Note that the CRs'
gyroradius does not enter the resonance condition. Hence the CRs that
are in resonance with compressible waves experience Fermi-II
acceleration irrespective of the length scale of the perturbation.

However, the resonance changes the component of the particle momentum
parallel to the mean magnetic field, which over time leads to
increasing anisotropy in the particle distribution that decreases the
efficiency of reacceleration with time. As in \citet{brunetti11}, we
assume that there exists a mechanism---such as the firehose
instability---that isotropises the CR distribution function at the
gyroscale and on the reacceleration time scale, which ensures
sustained efficient reacceleration with time.

All of the physics of turbulent reacceleration is effectively encapsulated in the diffusion coefficient \citep{brunetti07}, which can be rewritten as \citep{miniati15}:
\begin{equation}
D_{pp}(p) = \frac{ p^{2} \upi I_{\theta}(c_{s}/c)}{8 c} \langle k \rangle_{\mathcal{W}} \langle (\delta \varv_{\rm c})^{2} \rangle
\label{eqn:diffusion}
\end{equation}
where $I_{\theta}$ averages interaction rates over the CR pitch angle $\theta$; $I_{\theta} \approx 5$ for ICM conditions, and
\begin{equation}
\langle k \rangle_{\mathcal{W}} = \frac{1}{\langle (\delta \varv_{\rm c})^{2} \rangle} \int_{k_L}^{k_{\rm cut}} d k \, k \, \mathcal{W}(k) \approx \frac{s-1}{2-s} k_L \left( \frac{k_{\rm cut}}{k_L} \right)^{2-s} 
\label{eqn:k_W} 
\end{equation}
is an energy-averaged wavenumber, $k_L, k_{\rm cut}$ are the wavenumbers associated with the outer scale $L$ and the cutoff scale respectively, and we have assumed a total energy spectrum (composed of both kinetic and potential energy, where the two are assumed to be in equipartition, \citealt{sarkar11}): 
\begin{equation}
\mathcal{W}(k) = \frac{(s-1) \langle (\delta \varv_{\rm c})^{2} \rangle}{k_L} \left( \frac{k}{k_L} \right)^{-s} 
\label{eqn:Wk}
\end{equation}
which defines the normalisation $\langle (\delta \varv_{\rm c})^{2} \rangle$ (the subscript 'c' emphasizes that we specialise to compressive modes). Intuitively, we can understand the form of the diffusion coefficient from the fact that for second-order Fermi acceleration, $\dot{p} \sim \Delta p/\tau \sim p\,\langle k c \rangle_{\mathcal{W}} (\delta \varv_{\rm c}/c)^{2}$, where $\tau^{-1} \sim \langle k c \rangle_{\mathcal{W}}$ is the energy averaged wave-particle interaction rate, and $\Delta p \sim p (\delta \varv_{\rm c}/c)^{2}$ is the typical momentum change during wave-particle scattering. Thus: 
\begin{equation}
  \label{eqn:Dpp}
D_{pp} \sim p \dot{p} \sim p^{2} \langle k c \rangle_{\mathcal{W}} \left( \frac{\delta \varv_{\rm c}}{c} \right)^{2}. 
\end{equation}

Equations (\ref{eqn:diffusion}) and (\ref{eqn:Wk}) make the important aspects of turbulence clear: the energy in compressive modes $\langle (\delta \varv_{\rm c})^{2} \rangle$, the inner and outer scale ($k_{\rm cut}$ and $k_L$), and the slope of the energy spectrum $s$. All of these can vary spatially and temporally. 

We adopt a Kraichnan spectrum ($s=3/2$) for the fast modes, as seen in simulations \citep{cho03}. This can also be written as: 
\begin{equation}
  \label{eq:Wk}
  \mathcal{W}(k) =
\sqrt{2/7\,I_L\,\rho\,\langle \Vph \rangle}\,k^{-3/2},
\end{equation}
where $I_L$ is the volumetric injection rate of turbulence at scale $L$ (which
is assumed to be constant), $\varv_{\rm ph}\approx c_{\rm s}$ is the phase speed
of waves. Since the Kraichnan spectrum is critical in what follows, it is
  worthwhile taking a moment to recount the origin of the spectral slope $s$ and
  the cascade rate. In the magnetically dominated regime, two counter-propagating
  wave packets of scale $l$ interact on a wave crossing time $\tau_{\rm
    ph}=l/\varv_{\rm ph}$, rather than the eddy turnover time $\tau_{\rm edd} =
  l/\varv_{\rm l}$.  Since $\tau_{\rm ph} \ll \tau_{\rm edd}$, each interaction
  results in a small velocity change $\delta \varv_{\rm l} \sim \varv_{\rm l}
  (\tau_{\rm ph}/\tau_{\rm edd})$. If these changes behave like a random walk,
  the cascade time $\tau_{\rm l}$ it takes for an eddy to become non-linear
  ($\Delta \varv \sim \varv_{\rm l})$ and cascades to smaller scales occurs with
  a characteristic velocity $\varv_{\rm l} \sim \delta \varv_{\rm l} (\tau_{\rm
    l}/\tau_{\rm ph})^{1/2} \sim \varv_{\rm l} (\tau_{\rm ph}/\tau_{\rm edd})\,
  (\tau_{\rm l}/\tau_{\rm ph})^{1/2}$, which we can solve to obtain the cascade
  time:
\begin{equation}
  \label{eqn:tau_l}
\tau_{\rm l} \sim \left(\frac{\tau_{\rm edd}}{\tau_{\rm ph}}\right)^{2} \tau_{\rm ph} \sim \frac{l \varv_{\rm ph}}{\varv_{\rm l}^{2}} 
\end{equation}
where we have implicitly assumed isotropy. This implies a dissipation rate $\epsilon \sim \varv_{\rm l}^{2}/\tau_{\rm l} \sim \varv_{\rm l}^{4}/(l \varv_{\rm ph})$, or $\varv_{\rm l} \sim (\epsilon l \varv_{\rm ph})^{1/4}$. Thus, since $k E(k)\sim \varv_{\rm l}^{2}$, this gives a kinetic energy spectrum: 
\begin{equation}
E(k) \sim (\epsilon \varv_{\rm ph})^{1/2} k^{-3/2}.
\end{equation}
 
It is important to realise that this Kraichnan scaling only applies at small scales, where turbulence is magneto-hydrodynamical. At large scales, turbulent motions are subsonic (${\mathcal M}_{\rm s} \sim 0.2-0.5$) but super-Alfv{\'e}nic (${\mathcal M}_{\rm A} \sim 5$). Thus, except in the case where motions are transsonic and weak shocks become important, motions are fundamentally hydrodynamic and turbulence follows a Kolmogorov ($s=5/3$) spectrum. However, while the turbulent energy density $\eps_{\rm turb} \propto l^{2/3}$ decreases at small scales, the magnetic energy density (which is dominated by the large scale mean field) is scale-independent. Thus, at some scale $l_{\rm A} \sim l_L {\mathcal M}_{\rm A}^{-3}$, where $\eps_{\rm turb} \sim \eps_{B}$, the magnetic field becomes dynamically important, and turbulence transitions to the MHD regime. The cutoff scale in equation (\ref{eqn:Wk}) can be computed by setting the turbulent cascade rate for fast modes to the transit time damping rate on thermal electrons. This yields \citep{brunetti07,miniati15}:
\begin{equation}
k_{\rm cut} \approx A k_{\rmn{outer}} \frac{\langle (\delta \varv_{\rmn{outer}})^{2}\rangle^{2}}{c_{\rm s}^{4}}
\label{eqn:k_c1} 
\end{equation}
where $A \approx 11 000$. If there is an unimpeded cascade from the injection to the cutoff scale, $k_{\rmn{outer}}=k_L$ and $\varv_{\rmn{outer}}=\varv_{\rmn{c}}$. However, in our case the hydrodynamic (Kolmogorov) cascade transitions to the MHD cascade at the Alfv{\'e}n scale so that $k_{\rmn{outer}}=k_{\rmn{A}}$ and $\varv_{\rmn{outer}}=\varv_{\rmn{A}}$ and hence, we get\footnote{Note that this differs from previous work, which assumes Kraichnan turbulence from the injection scale onward and effectively adopts wave number and compressible velocity at the outer scale $L$ for this estimate of $k_{\rm cut}$.}
\begin{equation}
k_{\rm cut} \approx A k_{\rmn{A}} \beta^{-2}\,.
\label{eqn:k_c} 
\end{equation}

This gives $2\upi/k_\rmn{cut}\sim 0.1-1$~kpc in the ICM. This constitutes an
effective mean free path for CRs, unless plasma instabilities can
mediate interactions between turbulence and particles on smaller
scales \citep{brunetti11}, a possibility we discuss in Section~\ref{sect:self-limiting}. Another possibility is that compressible modes dissipate in weak shocks, resulting in Burgers' turbulence $s=2$ \citep{kowal10,porter15,miniati15}. If Burgers turbulence dominates, then particle acceleration rates are too slow in the face of cooling processes to explain radio haloes \citep{miniati15}, and an alternative model for radio haloes is required. 

The spatial profile of injected turbulence depends on the details of the
merger such as time during the merger, the impact parameter, the
merger mass ratio, and the degree of cluster anisotropy
\citep{miniati15}. We parametrize these uncertainties and
assume that volumetric injection rate of turbulent energy,
$I_L\propto \eps_\rmn{th}^{\alpha_\rmn{tu}}$, and determine the
normalisation by requiring that the turbulent energy in compressible
modes $E_\rmn{turb}=\int \int \mathcal{W}(k) \rmn{d}k \rmn{d}V =
X_\rmn{tu} E_\rmn{th}$, where $E_\rmn{th}$ is the total thermal
energy. Given these definitions, one can show that: 
\begin{equation}
  \label{eq:Dpp_scaling}
  D_{pp}\propto \frac{I_L}{\rho c_{\rmn{s}}} \propto 
X_\rmn{tu}^2k_L \eps_\rmn{th}^{\alpha_\rmn{tu}-1}\sqrt{T}\,.
\end{equation}

What is a typical energy density in compressive fast modes? The total turbulent
energy is typically $\sim 15-70$ per cent of the thermal energy in a cluster
\citep{vazza11}; it rises rapidly during a merger. The compressible component is
$\sim 20-40$ per cent of the total turbulent energy, and shows more rapid temporal
variations compared to the incompressible component
\citep{2013ApJ...771..131B,miniati15}. Note the compressible component in
cluster simulations is much larger than in stirring box simulations
with similar Mach numbers \citep{kowal10,lynn14}. This is likely due to the
compressive nature of turbulent driving (transsonic infall and merger), whereas
idealised simulations tend to use incompressible solenoidal driving and allow
compressive fluctuations to develop on their own. Overall, we adopt a
compressive energy density which is $X_{\rm tu} \sim 0.2$ of the thermal energy
as a baseline estimate.

The important effects are best summarised in terms of the acceleration rate
  which is governed by advection in momentum space:
\begin{equation}
  \tau_{\rm D}^{-1}\equiv \Gamma_{\rm D}\equiv
  \frac{\dot{p}}{p}=p^{-3}\frac{\partial}{\partial p}\left(p^2D_{pp}\right)
  =\frac{4 D_{pp}}{p^{2}}.
\end{equation}
In the last step we have used that $D_{pp}\propto p^2$ for turbulent
reacceleration (equation \ref{eqn:Dpp}). Hence the acceleration time is independent of momentum.
This should be compared against the lifetime of turbulence, $\tau_{\rm cl}$, and the cooling time $\tau_{\rm cool}$. In Table~\ref{tab:timescales}, we show both the thermal quantities and the
timescales for CR cooling and (re)acceleration for three different
spatial regions of the RH. The densities \citep{1992A&A...259L..31B} and temperatures \citep{2009ApJ...696.1886B,2001A&A...365L..67A} are derived from X-ray observations. To calculate synchrotron cooling times, we use $B$-fields derived from Faraday rotation measurements \citep{bonafede10}. To calculate the acceleration time, we need to assume an outer scale. We assume an injection scale $k_L=2\upi/\lambda_L$, where $\lambda_L = 100$ kpc, which were assumptions adopted in previous work \citep{2006MNRAS.366.1437S,brunetti07, brunetti11}. This length scale corresponds to an eddy turnover time on the outer scale of $2\upi\lambda_L\varv_L^{-1} \sim 1.2$~Gyr if $\varv_L \sim 500 {\rm km \, s^{-1}}$, as is characteristic of a merger. Note that hydrodynamical simulations of clusters have sometimes found larger $\lambda_L$, in some cases comparable to the size of the cluster \citep[e.g. $\lambda_L \sim 1 {\rm Mpc}$ in ][]{vazza11,miniati15}. This choice is degenerate with $X_{\rm tu}$; in Section~\ref{sect:self-limiting} we argue that $k_{\rm A}$ is a more appropriate choice, but also find that when the decay time is appropriately scaled, we are relatively insensitive to the choice of $k_L$. We present $\tau_{\rm D}$ for 3 different models, which we discuss in the next section. The reacceleration timescale
$\tau_{\rm D}$ is similar between our three models, where the
difference comes from turbulent profile parametrized with
$\alpha_\rmn{tu}$. This implies that even small differences in the
turbulent profile could impact the seed CRs significantly. Finally, we adopt an duration of acceleration $\tau_{\rm cl} = 650$ Myr, in line with previous assumptions \citep{brunetti07}, roughly corresponding to the turbulent decay time. 
\begin{table}
  \caption{Thermal quantities and timescales for different spatial
    regions in a Coma like cluster.}
\begin{tabular}{l c  c c}
\hline
\hline
& & spatial regions & \\
 & $0.1\,\RH^{(2)}$ & $0.3\,\RH^{(2)}$ & $\RH^{(2)}$   \\
\hline
thermal quantities$^{(1)}$ & & & \\
$\rho$ [$10^{-27}$ g cm$^{-3}$] & 3.0 & 1.6 & 0.15 \\
$T$ [$10^{8}$ K] & 1.4 & 1.0 & 0.58 \\
\hline
timescales$^{(1)}$ & & & \\
$\tau_\rmn{D}$(\Mprimary)$^{(3)}$ [Gyr] & 0.45 & 0.44 & 0.39 \\
$\tau_\rmn{D}$(\Mstream)$^{(3)}$  [Gyr] & 0.50 & 0.47 & 0.34 \\
$\tau_\rmn{D}$(\Mflatturb)$^{(3)}$  [Gyr] & 0.69 & 0.56 & 0.27 \\
$\tau_\rmn{IC/sync}(P=10^4\,m_{\rmn{e}}c)^{(4)}$ [Gyr] & 0.11 & 0.15 & 0.22 \\
$\tau_\rmn{had}(P=100\,m_{\rmn{p}}c)^{(5)}$ [Gyr] & 2.4 & 4.5 & 47 \\
$\tau_\rmn{Coul}(P=\,m_{\rmn{e}}c)^{(6)}$  [Gyr] & 0.0092 & 0.017 & 0.17 \\
\hline
\end{tabular}
\begin{quote}
 Notes: \\ 
 (1) Median quantities from our simulated post-merging cluster g72a derived 
 during last 300~Myrs in time. \\
 (2) Radius of the giant radio halo in Coma where $\RH\approx0.6\,R_{200}$.\\
 (3) Fermi-II reacceleration for both electrons and protons at all energies.\\
 (4) Inverse Compton and synchrotron cooling for electrons.\\
 (5) Catastrophic losses for protons.\\
 (6) Coulomb cooling for electrons (protons factor $m_{\rmn{e}}/m_{\rmn{p}}$ smaller).\\
 \label{tab:timescales}
  \end{quote}
\end{table}

\section{Cosmological simulations}
\label{sec:results}

In this section, we solve the Fokker-Planck equation for CR transport on
Lagrangian particle trajectories through cosmic history. We then use our
parametrized model of turbulence to apply turbulent re-acceleration, and
compare radio halo profiles against observations of the Coma cluster. We follow
the philosophy of adopting assumptions roughly in line with previous successful
models, which are now confronted with more accurate calculations of the seed CR
population, and find the minimal modifications required to fit observations. We
re-examine these choices in Section~\ref{sect:param_comp}.

Since the standard vanilla model requires a CR seed population which is
inconsistent with the simulations (Figure~\ref{fig:Edens}), some modifications are
necessary, either in the seed CR population or in the properties of
turbulence. We focus on three scenarios. (i) In model {\em M-primaries}, we
assume that CR electrons, which have been accelerated by cosmic formation shocks
and successively cooled by inverse Compton and synchrotron losses, form a fossil
seed population for reacceleration. (ii) In model {\em M-streaming}, we account
for the outward streaming of central CRps, which produces a flat CR distribution
in the ICM and equivalently a flat secondary seed population of CRes for
reacceleration. (iii) In model {\em M-turbulence} we adopt a spatially flatter
turbulent profile than what was adopted before but assume that seed CRps and
secondary CRes follow the steep profile that is suggested by structure formation
simulations.

\subsection{Modelling diffusive shock acceleration}
\label{sec:cosmo_sim}
In this paper we focus on our simulated cluster, g72a, which is a
massive cluster of mass $M_{200}=1.6\times10^{15}\,M_\odot$ that
experienced a merger about 1 Gyr ago
\citep{2009MNRAS.399..497D}. Since the cluster mass, density and
temperature profiles are all similar to the well studied Coma cluster
\citep{2007MNRAS.378..385P,pinzke10}, we will compare our calculations
to radio and gamma-ray observations of Coma.

We use a simple test-particle model for the CR acceleration and
injection, where each shock injects CRs that trace a power-law in
momentum,
\begin{equation}
  f_\p(p,t) = C(t)\,p^{\alpha_\rmn{inj}}\,,\quad
  \alpha_\rmn{inj}=\frac{(\gamma_\rmn{ad}+1)\mathcal{M}^2}
        {(\gamma_\rmn{ad}-1)\mathcal{M}^2+2}
\end{equation}
determined by the normalisation $C(t)$ and the spectral index
$\alpha_{\rmn{inj}}$ that depends on the adiabatic index
$\gamma_\rmn{ad} = 5/3$ and the Mach number of the shock
$\mathcal{M}$ \citep[see also ][]{1998ApJ...502..518Q,miniati01,pfrommer06}. It is given by the ratio of the upstream velocity
($\varv_2$) and the sound speed ($c_{\rm s}$). The CR number density and CR
energy density are derived from
\begin{eqnarray}
n_\CR &=&
\int_{p_\rmn{inj}}^\infty \dd p\, f_\p(p)\\
\eps_\CR &=&
\int_{p_\rmn{inj}}^\infty \dd p\, f_\p(p) \,E(p),
\end{eqnarray}
where $E(p) = (\sqrt{1+p^2} -1)\, m_{\rmn{p}}\,c^2$ is the kinetic energy of a
proton with momentum $p$. We adopt a fit to Monte Carlo simulations of the
thermal leakage process that relates the momentum of injected protons
($p_\rmn{inj}$) to the thermal energy ($p_\rmn{th}$) of the shocked plasma
\citep{kang11}:
\begin{eqnarray}
  \label{eq:qinj}
  p_\rmn{inj} &=& x_\rmn{inj} p_\rmn{th} =
  x_\rmn{inj} \sqrt{\frac{2 \,\kB T_2}{m_{\rmn{p}} c^2}}\,, \nonumber \\
  \rmn{where}\quad x_\rmn{inj} &\approx& 1.17 \frac{\varv_2}{p_{\rmn{th}}\,c} \left(1+
  \frac{1.07}{\eb}\right) \left(\frac{\mathcal{M}}{3}\right)^{0.1}\,.
\end{eqnarray}
Here $\eb = B_0/B_{\perp}$, $B_0$ is the amplitude of the downstream
MHD wave turbulence, and $B_{\perp}$ is the magnetic field along the
shock normal. The physical range of $\eb$ is quite uncertain due to
complex plasma interactions. In this paper, we adopt $\eb = 0.23$,
which -- as we will later see -- corresponds to a conservative maximum
energy acceleration efficiency for protons of $0.1$. To derive the
acceleration efficiency, $\zeta_\rmn{inj}$, we first have to infer the
particle injection efficiency, which is the fraction of downstream
thermal gas particles which experience diffusive shock acceleration
\citep[for details see][]{pinzke13},
\begin{equation}
  \label{eq:eta}
  \eta_\rmn{p,lin} =
  \frac{4}{\sqrt{\upi}}\,\frac{x_\rmn{inj}^3}{\alpha_\rmn{inj}-1}\,
  \rmn{e}^{-x_\rmn{inj}^2}.
\end{equation}
The particle injection efficiency is a strong function of
$x_\rmn{inj}$ that depends on both $\mathcal{M}$ and $\eb$. The
energy density of CRs that are injected and accelerated at the shock
(neglecting the CR back reaction on the shock) is given by
\begin{equation}
\label{eq:CR_energy} 
  \Delta\eps_\rmn{CR,lin} =
  \eta_\rmn{p,lin}(\mathcal{M})\,n_\rmn{th}(T_2)
  \,\frac{\eps_\CR}{n_\CR}
\end{equation}
and the CR energy injection and acceleration efficiency is:
\begin{equation}
  \zeta_\rmn{lin} =
  \frac{\Delta\eps_\rmn{\CR,lin}}{\Delta\eps_\rmn{diss}},
   \quad\mbox{where}\quad
  \Delta\eps_\rmn{diss} = \eps_\rmn{th2} - \eps_\rmn{th0}\,\left(\frac{\rho_2}{\rho_0}\right)^{\gamma_\rmn{ad}}\,.
\label{eqn:energy_frac}  
\end{equation}
The dissipated energy density in the downstream regime,
$\Delta\eps_\rmn{diss}$, is given by the difference of the thermal
energy densities in the pre- and post-shock regimes, corrected for the
adiabatic energy increase due to gas compression.

We limit the acceleration efficiency to $\zeta_\rmn{max}$ by
steepening the spectral index of the injected population
$\alpha_\rmn{inj}$ to $\alpha_\rmn{sub}$ so that $\zeta_\rmn{lin}
\leq \zeta_\rmn{max}$ is always fulfilled. The slope
$\alpha_\rmn{inj}$ impact $\zeta_\rmn{inj}$ via the mean energy per
particle, $\eps_\rmn{p}/n_\rmn{p}$. This
procedure conserves energy and is motivated by models of non-linear
shock acceleration where a sub-shock with a lower compression ratio
(and hence steeper spectral index) forms
\citep[e.g.,][]{2000ApJ...540..292E}. Given our assumed $\eb=0.23$, we
find that for strong shocks where $\alpha_{\rmn{inj}} \lesssim 2.3$ the spectral
slope is steepened by a maximum of $\sim 10$ per cent in low
temperature regimes ($\kB T\sim 0.1$~keV), while the steepening is much
smaller for high temperature regimes ($\kB T\sim 10$~keV) that are more
relevant for clusters. Since $p_\rmn{inj}$ remains fixed, so does the
CR number density $n_\CR$. Hence we can solve for the
renormalised normalisation constant $C_\rmn{sub}$ using $n_\CR$
and Eqn.~\ref{eq:eta}:
\begin{equation}
  \label{eq:Cp_sub}
  C_\rmn{sub}=\eta_\rmn{p,lin}\,(\alpha_\rmn{sub}-1)\,p_\rmn{inj}^{\alpha_\rmn{sub}-1}\,,
\end{equation}
where the new distribution function is given by $f_\p(p,t)= C_\rmn{sub}
p^{-\alpha_\rmn{sub}}$. We set an upper limit on the ratio of
accelerated proton-to-dissipated energy in the downstream of strong
shocks that varies from $\zeta_\rmn{max} \sim 0.01-0.1$, depending on
the adopted model (for more details, see section~\ref{sec:results}).

In our Galaxy, the CRe-to-CRp ratio at a few GeV is $K_{\rmn{ep}} \approx
10^{-2}$. Hence, we adopt this as a fiducial value for the CRe-to-CRp
acceleration efficiency in our models {\em M-streaming} and {\em M-turbulence}
\citep[see][for more discussion]{pinzke13}. However, as recent PIC simulations
have shown, this is likely very different at weak shocks, with electrons
efficiently accelerated at perpendicular shocks
\citep{2014ApJ...794..153G,2014ApJ...797...47G}, and ions (and electrons)
efficiently accelerated at parallel shocks \citep{2014ApJ...783...91C,
  Park2015}. Thus, depending on magnetic geometry, $K_{\rmn{ep}}$ could be
either larger or smaller. Some observations of radio relics suggest high values
of $K_{\rmn{ep}}$, due to the absence of gamma-ray emission, which probes the
CRp population \citep{2014MNRAS.437.2291V}. This suggests primary CRes as a
viable alternative scenario to secondary CRes as seeds for the giant RHs. In our
{\em M-primaries} scenario, the injected distribution of CRes is derived in the
same way as for the CRps. Once they have been accelerated to relativistic
energies, injected electrons and protons are indistinguishable. We therefore
assume that CRp and CRe have the same distribution function $f_\rmn{e}(p) =
K_\rmn{ep} f_\rmn{p}(p)$, with a different normalisation (due to differing
acceleration efficiencies) $K_\rmn{ep}=0.1$ \citep[which is viable for primarily
  perpendicular shocks,][]{2014ApJ...794..153G}. Given the unknown magnetic
geometry at cluster shocks, we investigate the consequences of this additional
degree of freedom.

\subsection{Radio emission profile}

\begin{figure*}
\begin{minipage}{1\columnwidth}
   \begin{center}\Large{\Mprimary:}\\
     \includegraphics[width=\columnwidth]{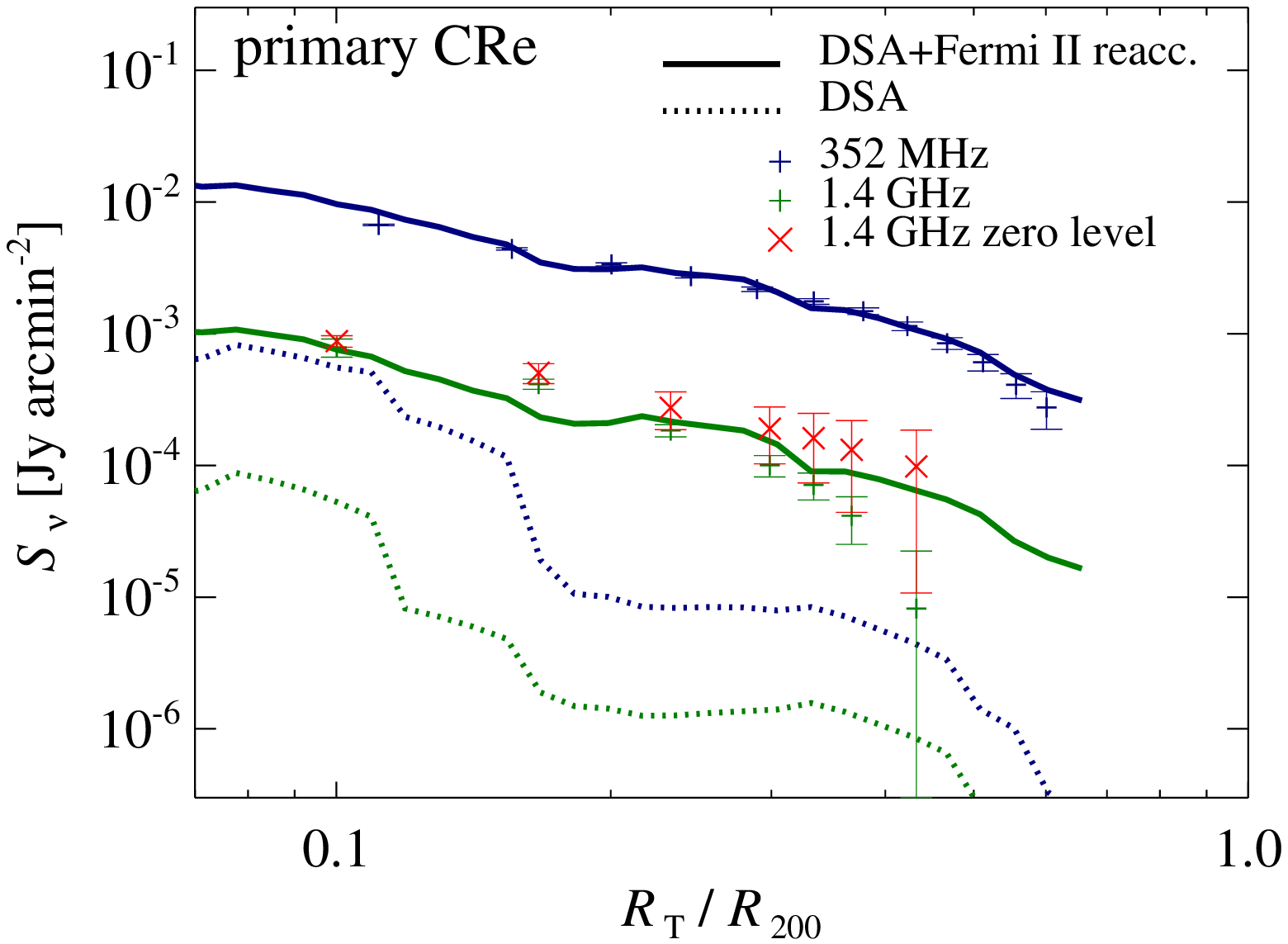}
   \end{center}
\end{minipage}
\begin{minipage}{1\columnwidth}
   \begin{center}\Large{\Mstream:}\\
     \includegraphics[width=\columnwidth]{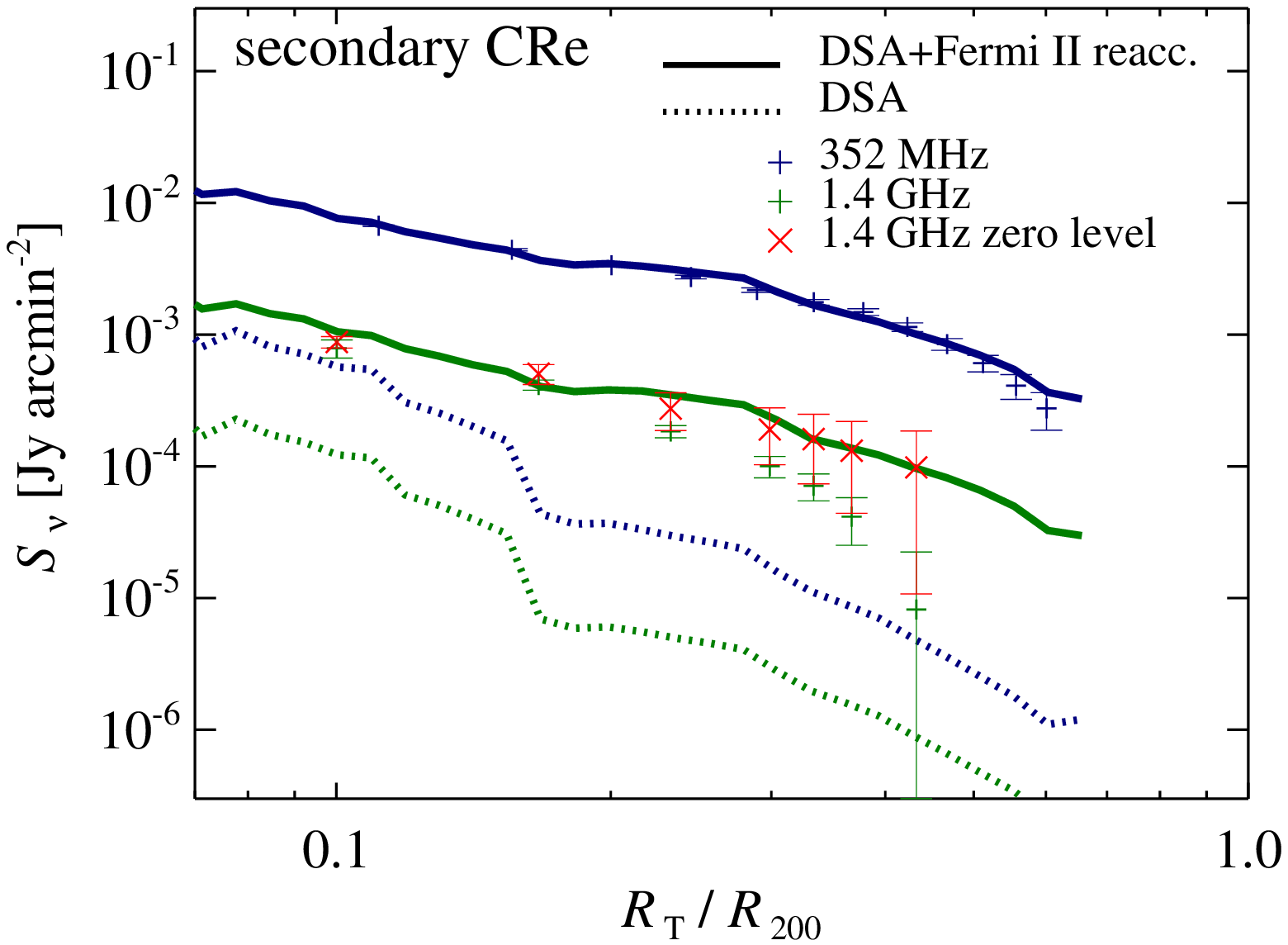}
   \end{center}
\end{minipage}
\\
\begin{minipage}{1\columnwidth}
  \begin{center}\Large{\Mflatturb:}\\ 
    \includegraphics[width=\columnwidth]{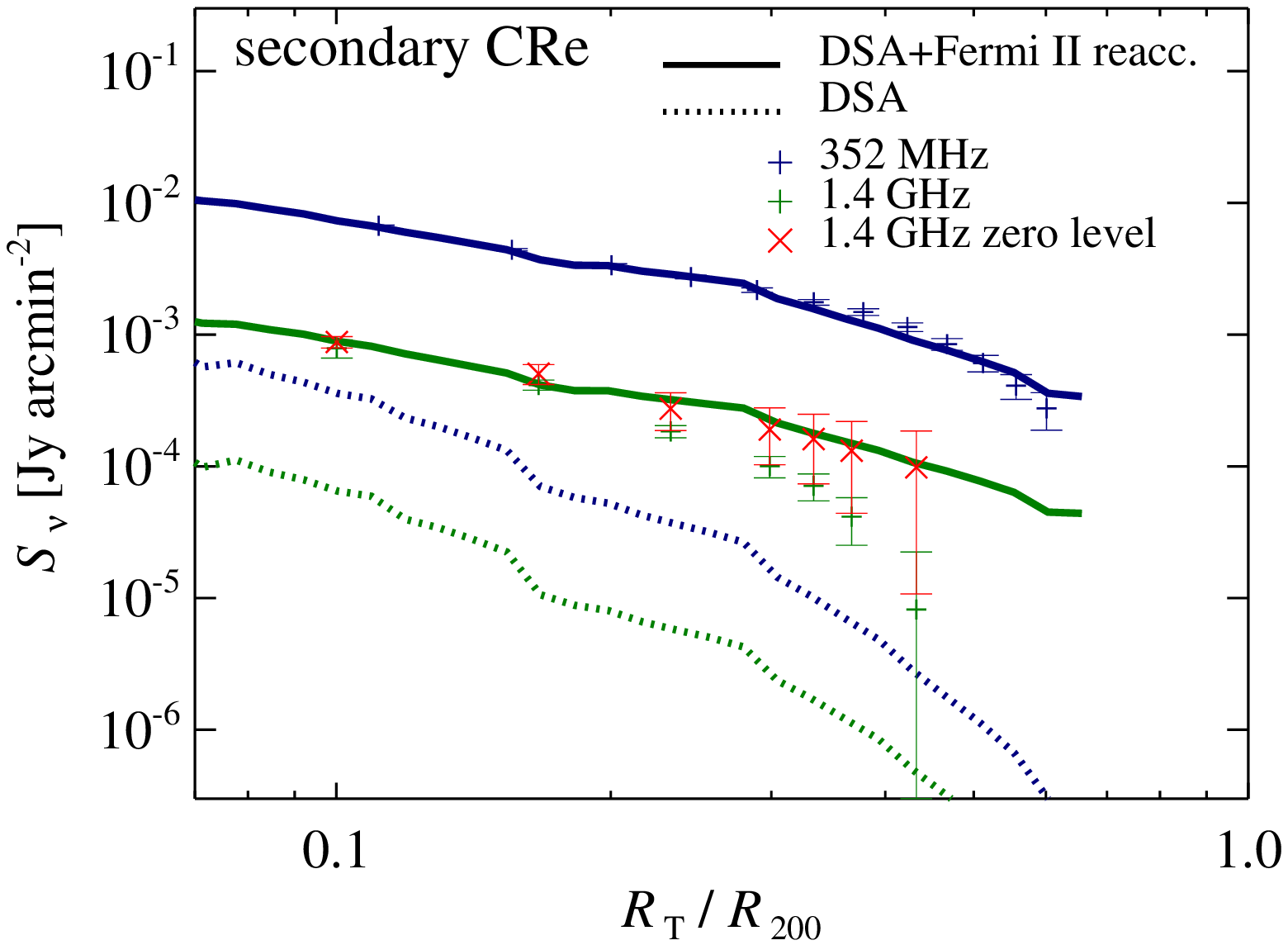}
  \end{center}
\end{minipage}
\begin{minipage}{1\columnwidth}
   \begin{center}\Large{\it Brunetti et al. (2012)}:\\
     \includegraphics[width=\columnwidth]{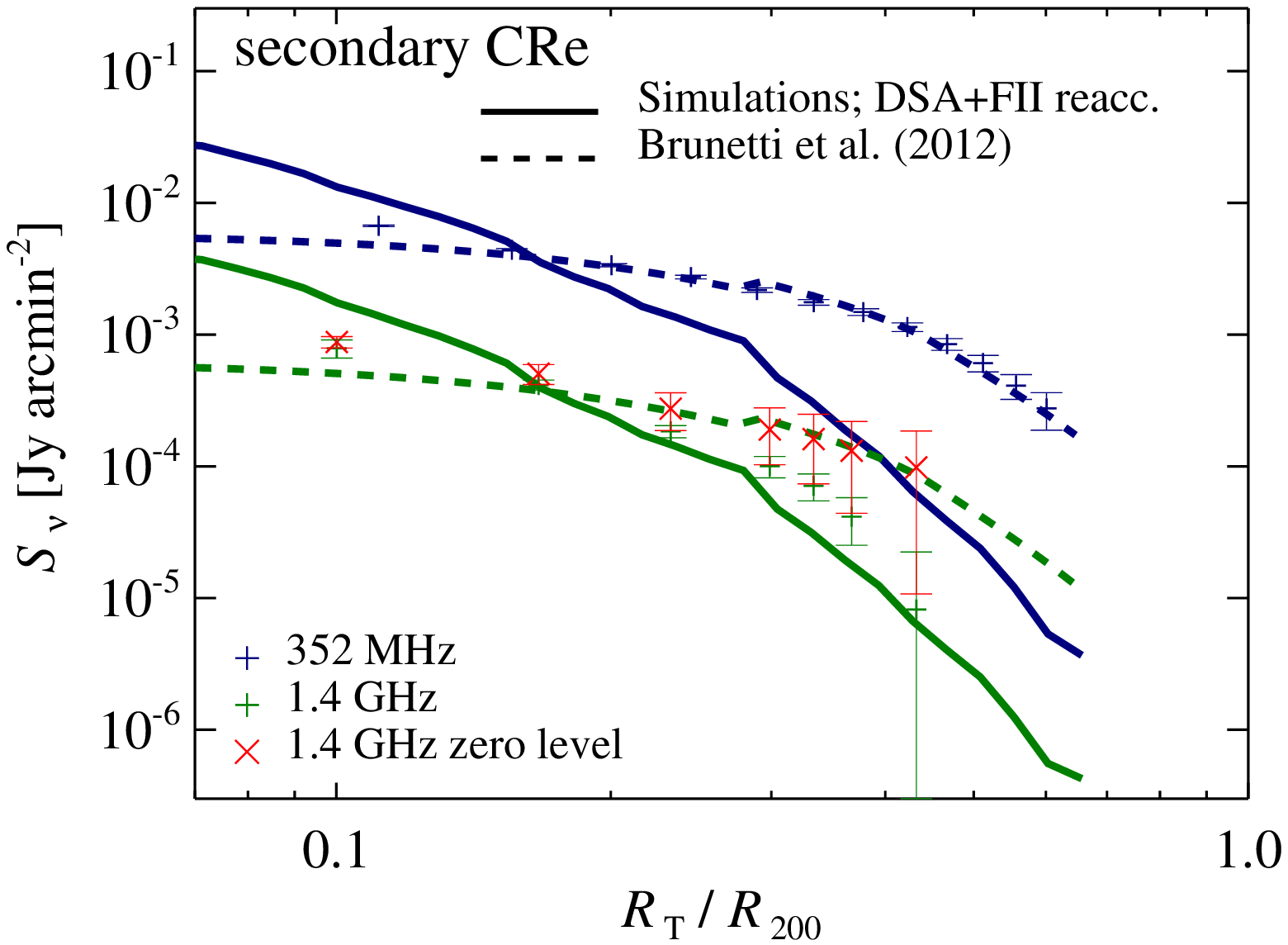}
   \end{center}
\end{minipage}
\caption{Radio surface brightness profiles of Fermi-II reaccelerated CR
  electrons of a simulated post-merging cluster similar to Coma. We compare
  profiles at 352~MHz \citep[blue lines and crosses,][]{brown11} to those at
  1.4~GHz \citep[green lines and crosses,][]{deiss97}. The red crosses show the
  reprocessed 1.4~GHz data, where a zero level of about 0.1 of the central value
  is adopted. The solid lines show predicted emission from a reaccelerated
  fossil population, while dotted lines show emission from a fossil population
  without reacceleration. The panels show the emission of our models \Mprimary
  (upper left panel), \Mstream (upper right panel), \Mflatturb (lower left
  panel), and simulated secondary electrons together with previous estimates
  \citep{brunetti12} for the Coma cluster (lower right panel).}
  \label{fig:sync_profile}
\end{figure*}

\begin{figure}
  \includegraphics[width=1.0\columnwidth]{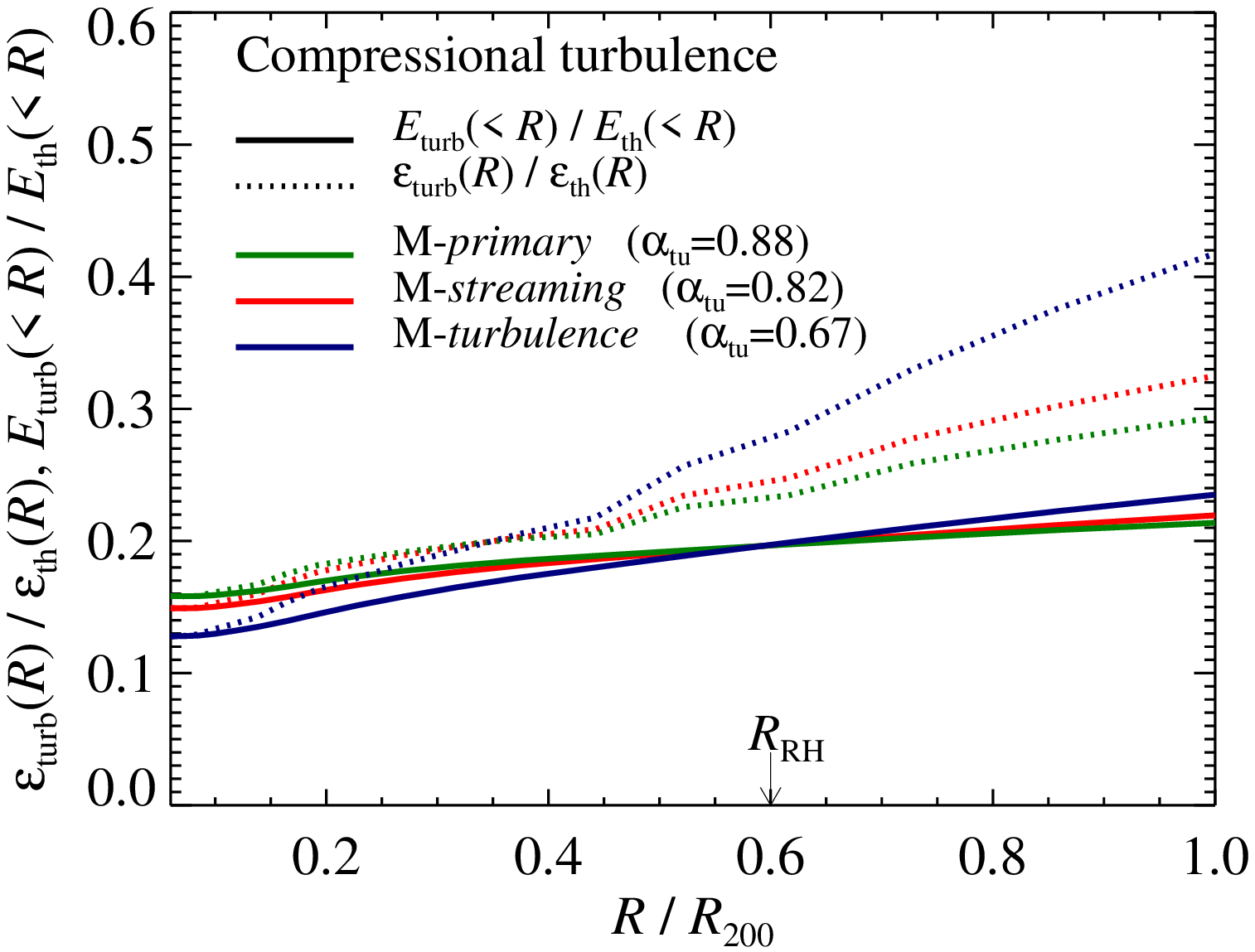}
  \caption{The ratio of turbulent-to-thermal energy densities (solid
    lines) and cumulative energies (dotted lines) in our three
    models. The energy densities are parametrized as
    $\eps_\rmn{turb} \propto
    \eps_\rmn{th}^{(\alpha_{\rmn{tu}}+1)/2}\,T^{-1/4}$ and
    normalised such that the total turbulent energy in compressible
    modes $E_\rmn{turb}$ for each scenario makes up about 20 per cent of the
    total thermal energy $E_\rmn{th}$ inside the radio halo
    ($R_\rmn{RH}\approx0.6R_{200}$). The turbulent profiles explore
    the uncertainty in the cluster turbulence and are motivated by the
    cosmological simulation in
    \citep{2009ApJ...705.1129L,2010ApJ...725.1452S,2011A&A...529A..17V}.}
  \label{fig:turb}
\end{figure}

In Figure~\ref{fig:sync_profile}, we show radial profiles for the radio
emission in all three scenarios in which the seeds undergo Fermi-II
reacceleration in turbulent fields that are shaped such as to
reproduce the Coma RH profile at 352~MHz.  Adopting our
parametrization for the volumetric injection rate of turbulent energy,
$I_L\propto \eps_\rmn{th}^{\alpha_\rmn{tu}}$, we find that to fit the observations, we require
$\alpha_\rmn{tu}= 0.67$ for \Mflatturb, $\alpha_\rmn{tu}= 0.82$ for
\Mstream, and $\alpha_\rmn{tu}= 0.88$ for \Mprimary. As a result, the
ratio of turbulent-to-thermal energy densities slightly increase with
radius as shown in Figure~\ref{fig:turb}.  The fine-tuning of these exponents is somewhat problematic, as we discuss in Section~\ref{sect:param_comp}. After turbulent
reacceleration, the volume-weighted, relative CRp energy density and
relative CRp number density inside the RH for \Mflatturb (\Mstream),
are found to be 2 (3) per cent and $2\times10^{-8}$ ($5\times10^{-8}$),
respectively.

Figure~\ref{fig:sync_profile} demonstrates that the modelled radio
profiles without turbulent reacceleration are too steep.  In the
bottom right panel of Figure~\ref{fig:sync_profile} (labelled with
Brunetti et al. 2012), we show that our simulated profiles of
reaccelerated CRs, which only take advective CR transport into
account, i.e. they neglect CR streaming or a flatter turbulent
profile, produce radio profiles that are too steep. Indeed, even using
the assumptions of previous work, where complete freedom in the seed
population was allowed, it is not possible to reproduce observations
in both frequencies in any model\footnote{Note that in previous work
  on the Coma cluster, $\eps_\rmn{turb} \propto \eps_\rmn{th}$ was
  adopted which approximately corresponds to $\alpha_\rmn{tu}= 1$
  \citep{brunetti12} and together with the different distributions for
  seed CRes constitute the main differences compared to our work.} -- with or without turbulent reacceleration.
Decreasing the acceleration efficiency with radius does not change
this conclusion much because of the weak radial dependence of
$D_{pp}(R)\propto \eps_\rmn{th}(R)^{\alpha_\rmn{tu}-1}
\sqrt{T(R)}$. This signals that the problem is generic and requires
either additional modifications to the plasma physics of acceleration
or a better understanding of potential observational systematics. In
addition there are differences in the simulated density and
temperature profiles in comparison to the observed profile in Coma
that impact the CR abundance as well as cooling and reacceleration.

\begin{figure}
  \includegraphics[width=1.0\columnwidth]{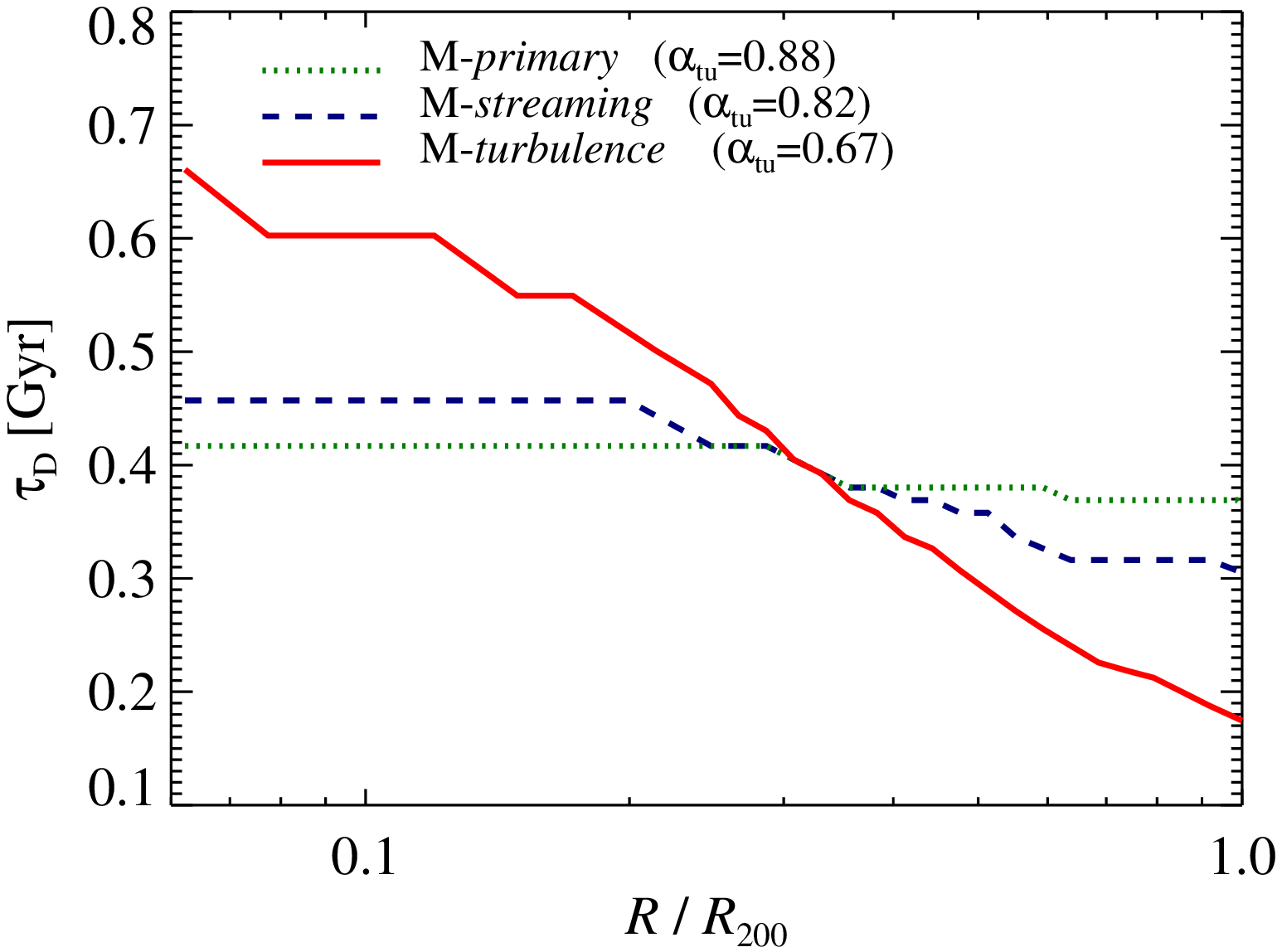}
  \caption{Turbulent reacceleration timescales for our simulated
    cluster g72a. We show in linear-log the reacceleration timescale
    ($\tau_{\rm D}$) as a function of radius $R$ for our three models:
    \Mprimary (green dotted line), \Mstream (blue dashed line), and
    \Mflatturb (red solid line). Note that the timescales are derived
    during the last 100~Myrs in time in our simulations.}
  \label{fig:tauD}
\end{figure}

In Figure~\ref{fig:tauD} we show how the turbulent reacceleration
timescales in our three models scale with radius. As expected, the
\Mprimary model with $\alpha_\rmn{tu}=0.88$ has the flattest profile
with $\tau_{\rm D}\approx0.4\,\rmn{Gyr}$, where the small dip at large
radius driven by the decrease in thermal energy density. The
\Mflatturb model has the flattest turbulent profile parametrized
by the smaller $\alpha_\rmn{tu}$ which explains the steepest $\tau_{\rm D}$
profile. Note that a fixed reacceleration timescale is required to
explain the observations at each radius and for each model (see also
Table~\ref{tab:timescales}). Since $\tau_\rmn{cl} \propto X_\rmn{tu}^2 k_L$,
these two parameters are degenerate and can be traded off one another.

In principle, reacceleration via TTD leads to spectral steepening with
particle energy due to the inefficiency of the acceleration process to
counter the stronger cooling losses with increasing energy. Since
synchrotron emission peaks at frequency $\nu_\rmn{syn}\simeq 1\,
B/\umu\rmn{G} (\gamma/10^4)^2\,\rmn{GHz}$, this translates into a
spectral steepening of the radio spectrum (see the left panel of
Figure~\ref{fig:sync_spectrum} where the continuous injection of
secondary CRes is absent). A given radio window samples higher energy
electrons for a decreasing field strength in the cluster
outskirts. Hence, the spectral steepening with energy should translate
into a radial spectral steepening \citep{brunetti12}. However, because
of the weak dependence of the electron Lorentz factor on emission
frequency ($\gamma\propto\sqrt{\nu_\rmn{syn}}$), this effect is only
visible in our simulations for $\nu_\rmn{syn}\gtrsim5$~GHz. Most
importantly, our simulated fluid elements at a given radius sample a
broad distribution of shock history, density and temperature, which
implies very similar synchrotron brightness profiles at
$\nu_\rmn{syn}=352$~MHz and 1.4 GHz. The discrepancy of the observed
and simulated 1.4 GHz profiles could instead be due to systematic flux
calibration error in single dish observations. These could arise, for
instance, due to errors in point source subtraction. Interestingly, we
can match the 1.4 GHz data if we reduce the zero point by adding 0.1
of the central flux to every data point; this flattens the outer
profile\footnote{Lawrence Rudnick, private communication.}.
Alternatively, this may point to weaknesses in the theoretical
modelling of the particle acceleration process and may require a
stronger cutoff in the particle energy spectrum.

\subsection{Radio spectrum}
\label{sect:radio_spec}
\begin{figure*}
  \includegraphics[width=1.0\columnwidth]{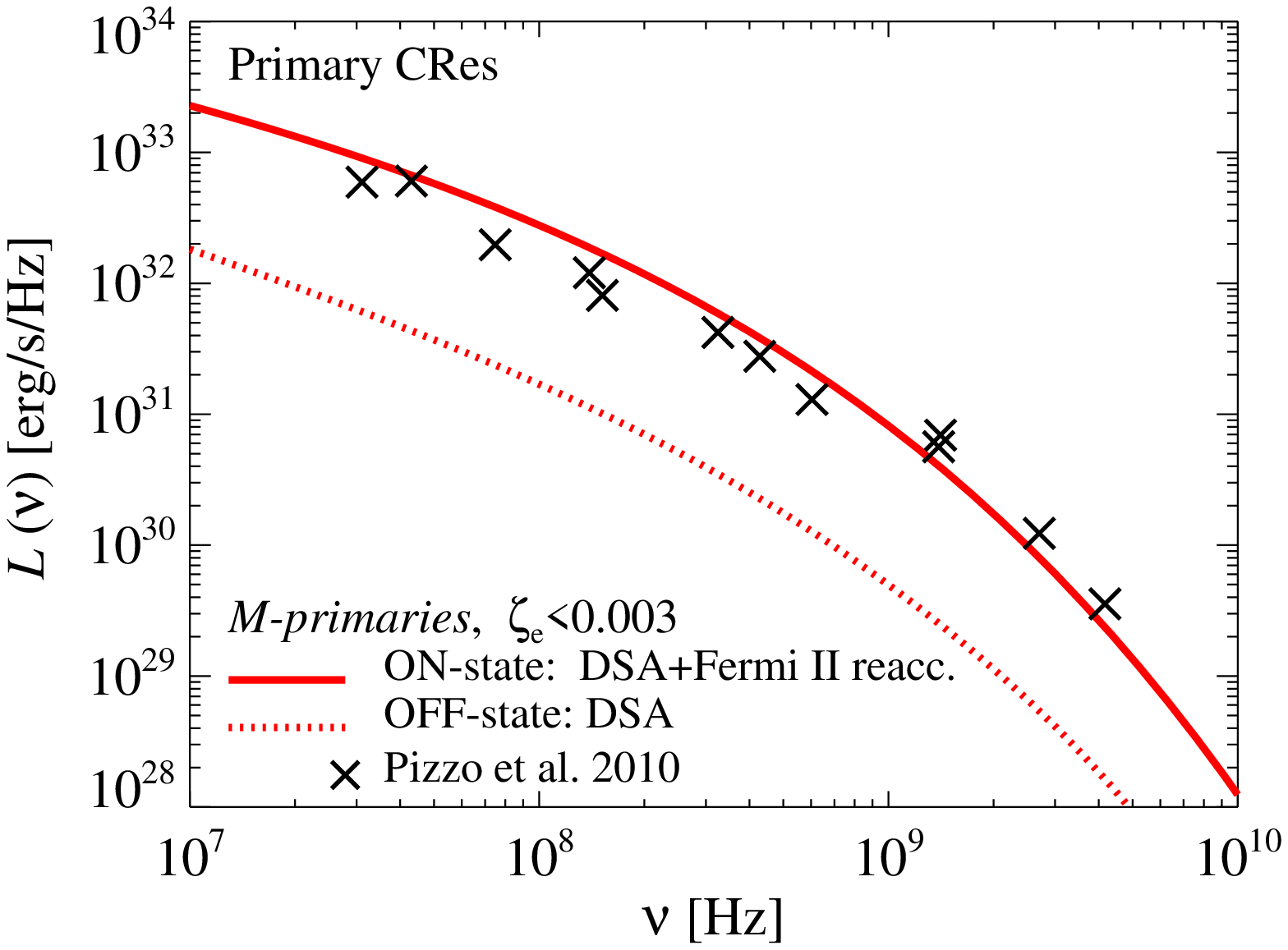}
  \includegraphics[width=1.0\columnwidth]{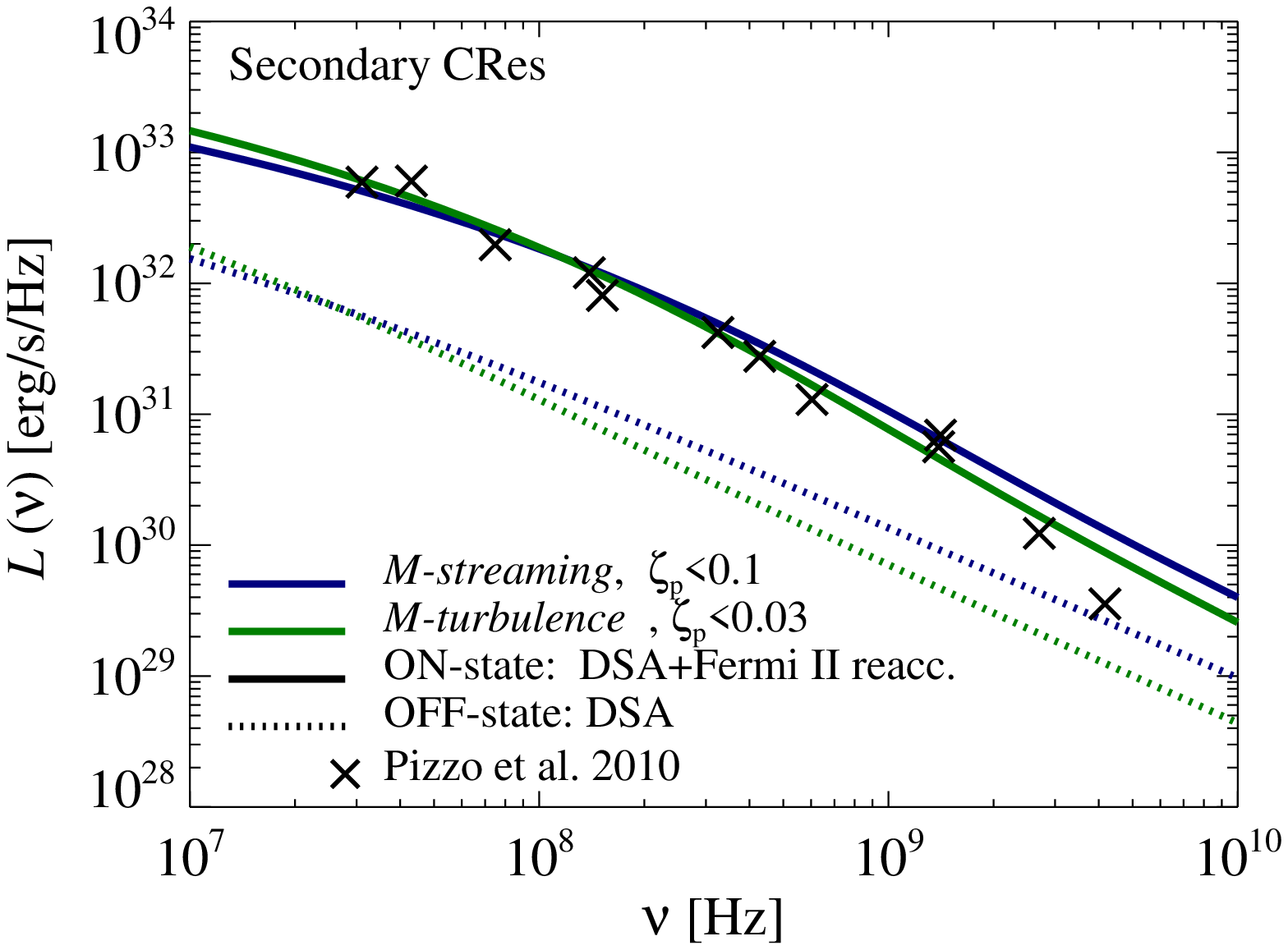}
  \caption{Radio synchrotron spectra. Lines are derived from
    simulations, while the black crosses are compiled from
    observations \citet{2010PhDT.......259P}. The solid lines show the
    DSA and reaccelerated CRs (On-state of the radio halo), while the
    dotted lines show CRs accelerated only by DSA (Off-state of the
    radio halo). The left figure shows the radio emission induced by
    primary CRes and the right figure shows the emission from
    secondary CRes. The different line colours represent our different
    models, \Mprimary (red line), \Mstream (blue line), and \Mflatturb
    (green line).}
  \label{fig:sync_spectrum}
\end{figure*}

In Figure~\ref{fig:sync_spectrum} we show that our three models that
include Fermi-II reacceleration can individually reproduce the
convexly curved total radio spectrum found in the Coma cluster. Seed
CRs in \Mstream and \Mflatturb that do not experience turbulent
reacceleration have a power-law spectrum in disagreement with
observations. In order to match both the spatial and spectral profiles
in Coma, we adopt an acceleration efficiency for the strongest shocks
in our three models \Mprimary, \Mstream, and \Mflatturb to
$\zeta_{\rmn{e}} <0.003$, $\zeta_{\rmn{p}} < 0.1$, and
$\zeta_{\rmn{p}}<0.03$, respectively. Following the Mach number
($\mathcal{M}$)-dependence of the acceleration efficiency suggested in
\cite{pinzke13}, the efficiency in weak shocks ($\mathcal{M}\sim
2.5-3.5$) that dominates the CR distribution function, has an
acceleration efficiency for protons $\zeta_{\rmn{p}}\sim0.0001-0.01$,
and for electrons $\zeta_{\rmn{e}}\sim 0.001$.

Interestingly, we find that the radio luminosity from clusters in the
OFF-state (DSA only) and ON-state (DSA and reacceleration) differ by
about a factor 10-20 in all our three models. This means that the
secondary CRes are dominated by the reaccelerated fossil CRes and not
from the CRes produced by reaccelerated CRps. However, for high
frequencies ($\nu_\rmn{syn}\gtrsim$ GHz) where synchrotron cooling is
more efficient than reacceleration, the emission is dominated by the
CRes produced in the continuous injection of electrons from
reaccelerated CRps. It is also worth mentioning that the radio
emission from secondary CRes are smoothly distributed around the
cluster because of the continuous injection, hence it is not
dominated by outliers.

However, for \Mprimary, the primary CRes that generate most of the
radio emission from the cluster in the OFF-state are dominated by only
a small fraction of the CRes. These electrons are injected very
recently and have not had time to cool yet. Hence we expect there to
be a large variance in the OFF-state of different simulated
clusters. As mentioned in section~\ref{sect:param_comp}, combining
radio observations with gamma-ray limits allows us to put a lower
limit to $X_\rmn{tu}$.  If $X_\rmn{tu}$ is smaller than in our adopted
models (where we assume $X_\rmn{tu}=0.2$), then the efficiency of DSA
has to be larger than $\zeta_\rmn{p}\sim 0.1$ for the secondary CRes
to reproduce the radio observations. However, since the turbulent
reacceleration acts on both the secondary CRes and the CRps, while
$\zeta_\rmn{p}$ only affects the CRps, \Mstream and \Mflatturb would
produce too much gamma-rays. Hence we conclude that
$X_\rmn{tu}\gtrsim0.2$ if all other parameters are kept
fixed. Although, we caution the reader to take this limit too
stringent because of the uncertainty in $k_L$ and $\tau_\rmn{cl}$ that
impact $X_\rmn{tu}$ for a fixed $\tau_\rmn{D}$. This parameter space
needs to be explored further in future work in order to put more
stringent limits on the level of turbulence in clusters using radio
and gamma-ray observations in combination with turbulent reaccelerated
CRs.

\subsection{Gamma rays}
The gamma-ray emission from CRps that produce decaying neutral pions could be
substantial if the CRps are reaccelerated efficiently enough, hence it is
interesting to estimate this emission for our models and compare to upper
limits. We follow the formalism outlined in \cite{1999APh....12..169B} (and
references therein) and calculate the gamma-ray emission numerically for our
three models. We predict the gamma-ray emission from \Mflatturb (\Mstream) with
$F_\gamma(>500\,\rmn{MeV})=4\times10^{-10} (5\times10^{-10}) \mathrm{ \,ph\,
  s}^{-1}\mathrm{cm}^{-2}$. The fluxes from these models are slightly larger
than in \cite{brunetti12}, where the differences comes from our steeper CRp
profiles in addition to the simulation based formalism we rely on that accounts
for both Coulomb and hadronic losses during the build up of the CR distribution
in contrast to the scaling relations adopted in their paper. Interestingly the
gamma-ray flux from both our scenarios are just below recent Fermi-LAT limits
derived from a gamma-ray profile similar to \Mflatturb\footnote{Fabio Zandanel,
  private communication; see also
  \citet{2014MNRAS.440..663Z,2014ApJ...787...18A}; this will be probed in the
  next few years by Fermi-LAT.} where
$F_\gamma(>500\,\rmn{MeV})<5.3\times10^{-10} \mathrm{ \,ph\,
  s}^{-1}\mathrm{cm}^{-2}$.  The spectral index of the CRp distribution is
relatively steep ($\alpha_{\rmn{p}}\sim2.6$) for the CRp energies $E \gtrsim
10$~GeV that are relevant for the injection of radio-emitting secondary
CRes. The steep spectrum is ultimately a consequence of the shock history of the
simulated cluster, with a weak dependence on our test particle model for Fermi-I
acceleration \citep{pinzke13}, where we steepen the spectral index to avoid
acceleration efficiencies above $\zeta_{\rmn{p}} \sim 0.1$.

\section{Parameter Space Exploration and Overcoming Fine Tuning}
\label{sect:param_comp}
In this paper we rely on several critical parameters describing
relatively unknown non-thermal physics in the ICM. Here we develop a
simplified framework for our reacceleration model of secondary
electrons. We will explore how radio emission depends on the
parameters describing the spatial profile of CR protons and
turbulence. Our fiducial model is meant to be
compared against the Coma cluster.

\subsection{Methodology}

As we have seen before, the most uncertain aspects of radio halo
emission models are the profile of compressible turbulence (which
determines the amount of Fermi II acceleration) and the distribution
of pre-existing CRs. Hence we vary parameters describing the amount of
energy contained in turbulence, $X_{\rm tu}$ (defined by $E_{\rm
  turb} = X_{\rm tu} E_{\rm th}$), the spatial profile of
turbulence as parametrized by $\alpha_{\rm tu}$ (defined by $I_L
\propto \eps_{\rm th}^{\alpha_{\rm tu}}$, where $I_L^{} \propto
\varv_L^{3} k_L^{}$ is the injection rate of turbulence), as well as
the spatial CR profile that we will parametrize by $\alpha_{\rm CR,
  spat}$ (see below).  We hold fixed thermal plasma properties
(temperature and density profiles), $B$-field profiles, total CR energy
content, and the turbulent outer scale (corresponding to a wavenumber
$k_L$). The CR energy content is suggested by our simulations
(observations only give an upper bound;
\citealt{2012ApJ...757..123A}). We focus on the uncertain CR
distribution rather than the overall normalisation, since the impact
of the latter (an overall linear scaling) is clear.

In order to quickly explore this parameter space, we solve the
Fokker-Planck equation in static spherical shells for injection,
reacceleration, and losses of the CRs, i.e., we neglect Lagrangian
evolution during re-acceleration.  This ignores the effect of
adiabatic compressive heating of the CRs, though this is generally
subdominant (e.g., see Figure~7 of \citealt{miniati15}). Once
the CRs have been reaccelerated for $\tau_\rmn{cl} = 650\,\rmn{Myr}$,
we calculate the resulting radio emission numerically using the
formalism outlined in \cite{1979rpa..book.....R} and compare the
emission profiles and spectra as we vary one parameter at a time
relative to our fiducial model.

We adopt both the density \citep{1992A&A...259L..31B} and temperature
profiles \citep{2009ApJ...696.1886B,2001A&A...365L..67A} derived from
X-ray observations of the Coma cluster,
\begin{eqnarray}
n_{\rmn{e}} &=& n_0\,\left[1+\left(R/R_{\rmn{c}}\right)^2\right]^{-1.125},\nonumber\\
k_{\rmn{B}} T &=& 8.25\,\rmn{keV} \left[1+\left(2R/R_{200}\right)^2\right]^{-0.32},
\end{eqnarray}
with $n_0=3.4\times10^{-3}\,\rmn{cm}^{-3}$.  The virial and core radii
of Coma are given by $R_{200} = 2.3\,\rmn{Mpc}$
\citep{2002ApJ...567..716R} and $R_{\rmn{c}} = 294\,\rmn{kpc}$,
respectively. In accordance with Faraday rotation measure
measurements, we assume $B(r)=B_{0} (n/n_{0})^{\eta}$, where
$B_{0}=4.8 \umu$G and $\eta=0.5$ \citep{bonafede10}.

The bulk of the CRps are injected by relatively low Mach number shocks and
parametrized by $f_\rmn{p,inj}(p) = C_\rmn{inj}\,p^{-\alpha_\rmn{inj}}$, where
$\alpha_\rmn{inj}\approx 2.5$ in our simulations. The CRps approximately trace
the thermal gas with $C_\rmn{inj} \propto \eps_\rmn{th}^{\alpha_\rmn{CR,spat}}$
\citep{pinzke10,2016MNRAS.459...70V}, where the normalisation is fixed by the
injection rate of CR energy in the last 650~Myrs. Our simulations show that the
CR energy approximately amounts to 0.03 per cent of the thermal energy inside
the virial radius, i.e.  $\int_0^{R_{200}}\eps_\rmn{CR,inj} \rmn{d}
V\left(\int_0^{R_{200}}\eps_\rmn{th} \rmn{d}V\right)^{-1} = 0.0003$. The
spectrum of the initial CRp distribution is determined by the steady state
between injection and cooling,
\begin{equation}
 f_\rmn{p,0} \propto \frac{\int_p^\infty f_\rmn{p,inj}(p') 
\rmn{d}p'}{\displaystyle\left|{{\rmn{d}p}\over{\rmn{d}t}}\right|_{\rm Coul}+\frac{p}{\tau_{\rm had}}}\,,
\end{equation}
where we fix the normalisation by requiring $\int_0^{R_{200}}\eps_\rmn{CR,0}
\rmn{d}V\left(\int_0^{R_{200}}\eps_\rmn{th} \rmn{d}V\right)^{-1} = 0.003$. Note
that the injected {\em CR energy} in the last 650 Myr is smaller by about a
factor of 10 in comparison to the cumulative CR energy injected over the entire
cosmological history of the cluster. However, since the injected {\em CR energy
  rate} averaged over the formation time of the cluster is similar to that
during a merger, the CRes injected during the merger are especially important
for the radio emission above 1 GHz since these CRs have not yet had time to
cool. Similarly, the initial CRe distribution is given by the steady state
between cooling (Coulomb, inverse Compton, and synchrotron) and injection of
secondary CRes from $f_\rmn{p,0}$.  The diffusion constant $D_{pp} \propto
X_{\rm tu}^{2} \eps_{\rm th}^{\alpha_{\rm tu}-1} \sqrt{T} k_L$ is calculated for
each radial bin. All parameters and assumptions are similar to what is used for
our simulated cluster (see section~\ref{sec:results}). Our fiducial model
assumes $X_{\rm tu} = 0.2$, $\alpha_{\rm tu} =0.8$, $\alpha_{\rm CR, spat}=1.0$,
and $k_L=2 \upi/\lambda_L$ where $\lambda_L=100$ kpc. We also assume a fixed
acceleration time of $\tau_{\rm cl} = 650 \, {\rm Myr}$.

\subsection{Results}

\begin{figure*}
\begin{minipage}{1\columnwidth}
   \begin{center}\Large{radio profiles}\\
     \includegraphics[width=\columnwidth]{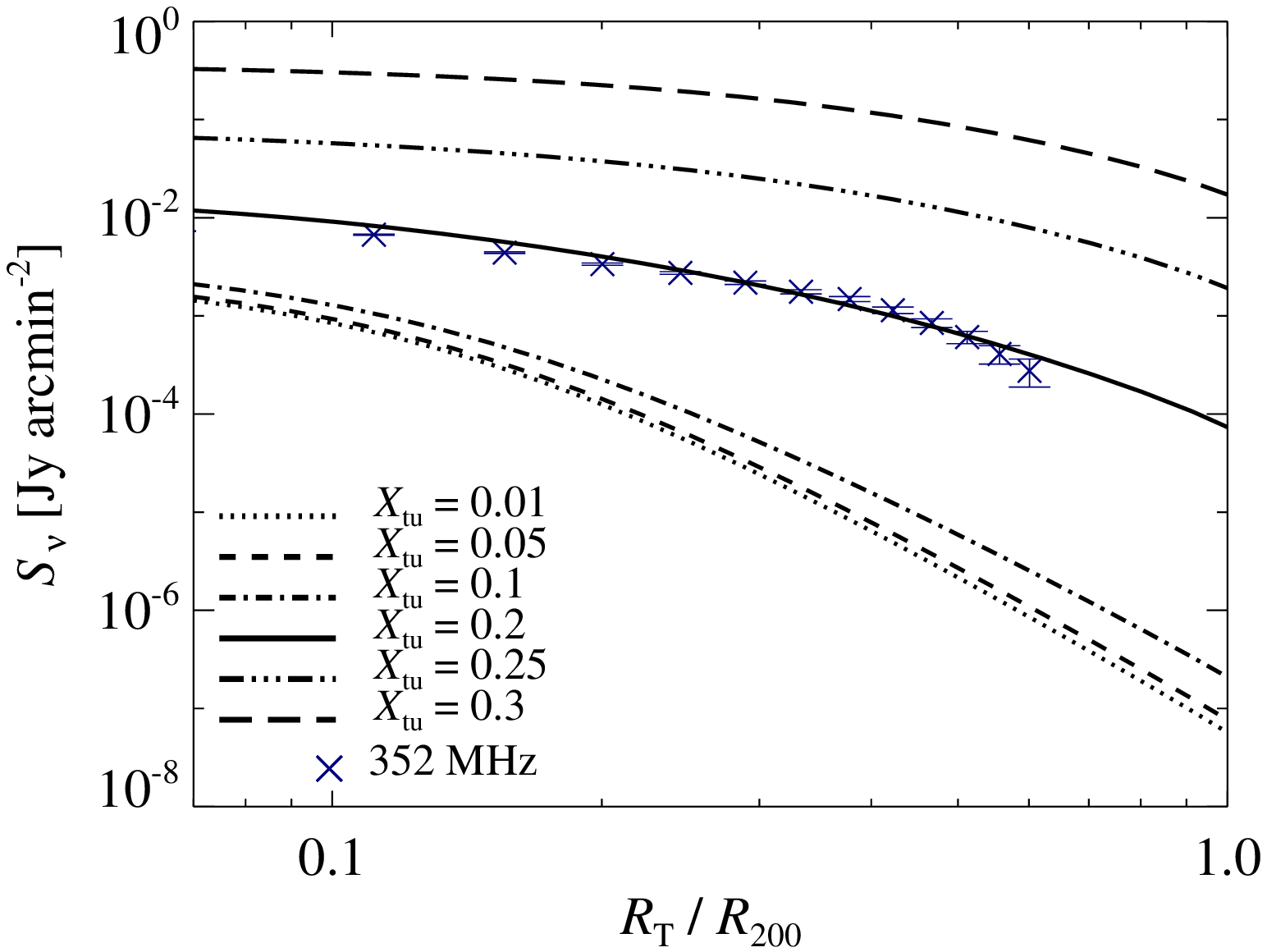}
   \end{center}
\end{minipage}
\begin{minipage}{1\columnwidth}
   \begin{center}\Large{radio spectra}\\
     \includegraphics[width=\columnwidth]{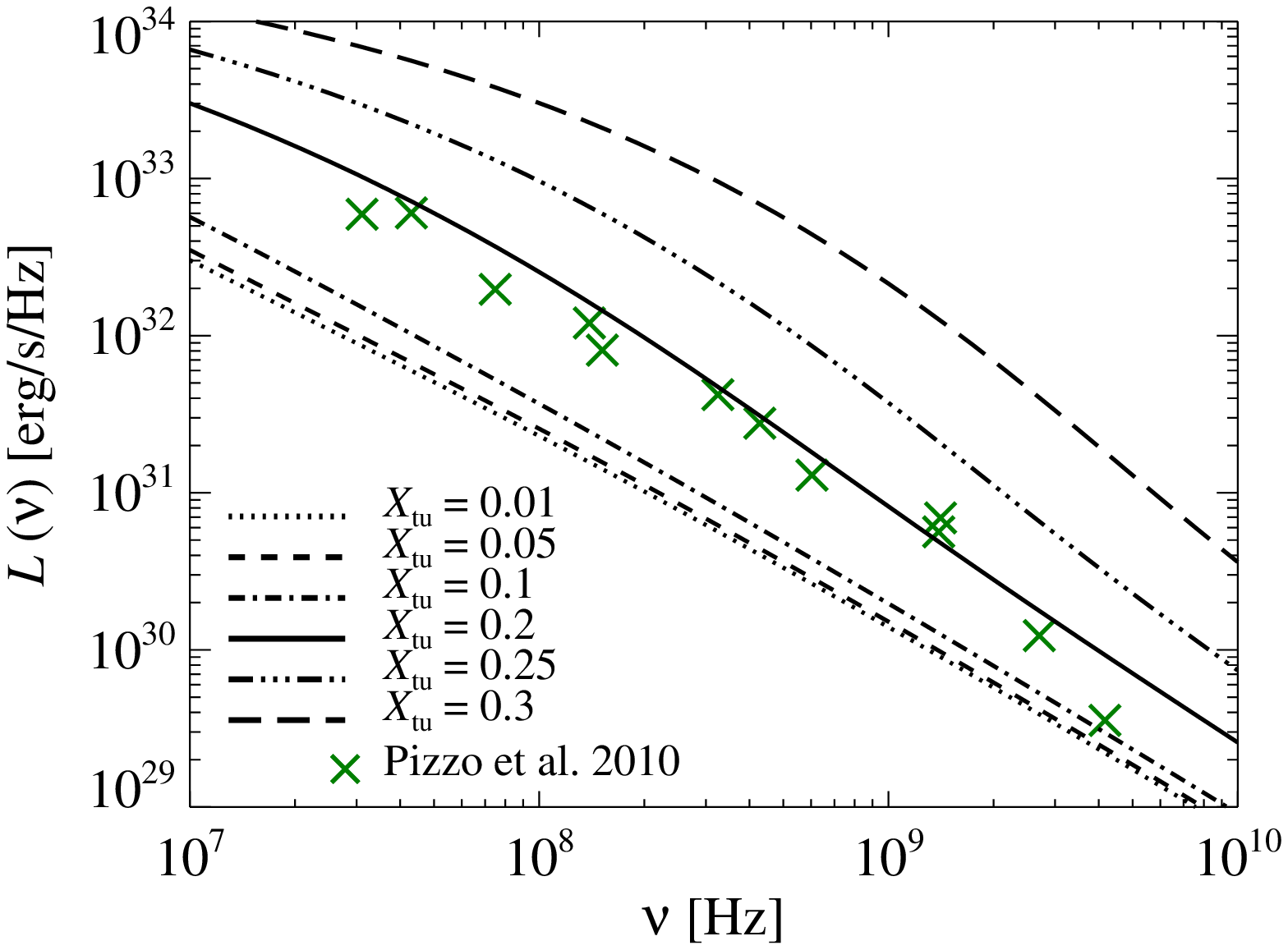}
   \end{center}
\end{minipage}
\\
\begin{minipage}{1\columnwidth}
  \begin{center}
    \includegraphics[width=\columnwidth]{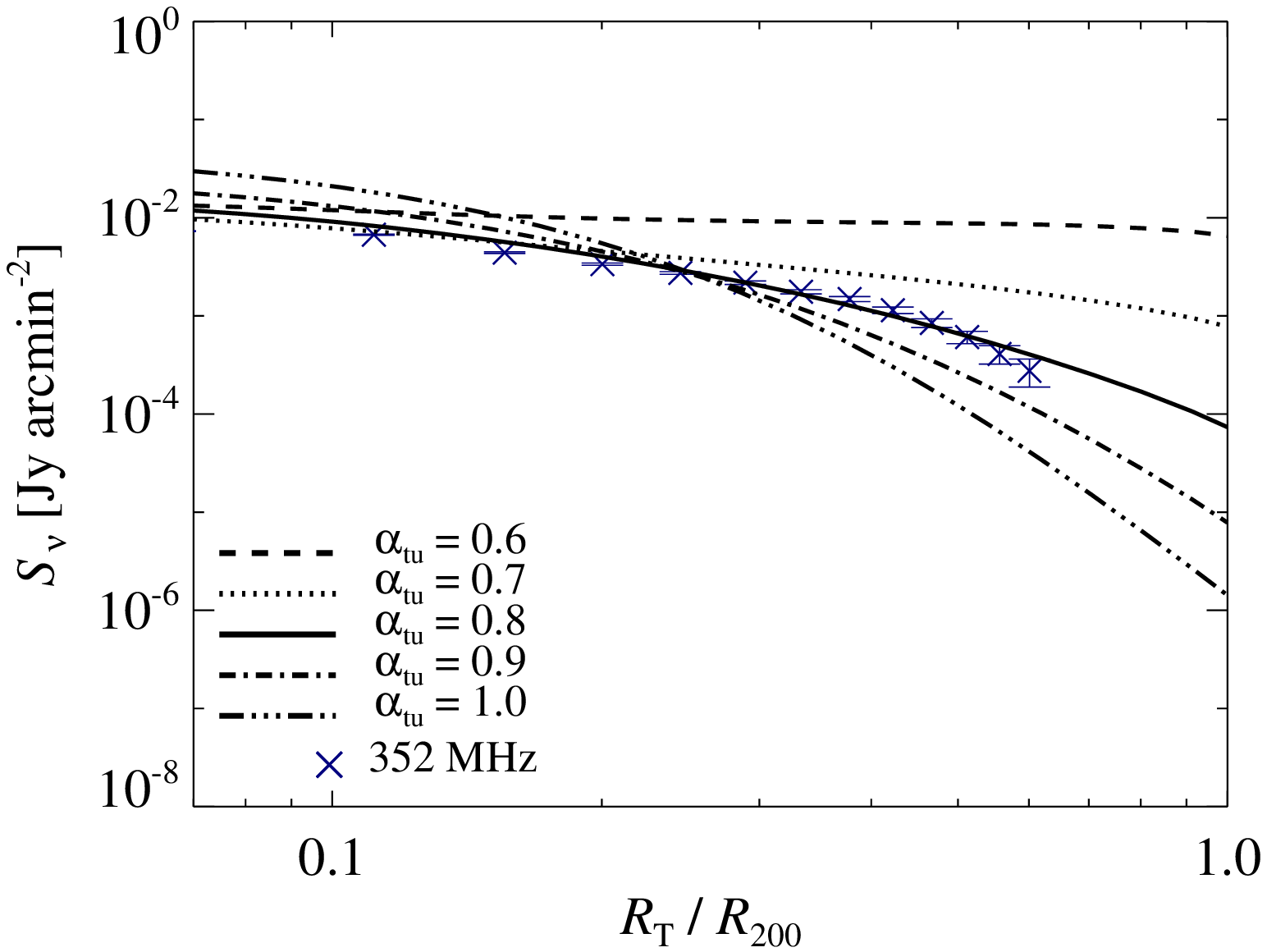}
  \end{center}
\end{minipage}
\begin{minipage}{1\columnwidth}
   \begin{center}
     \includegraphics[width=\columnwidth]{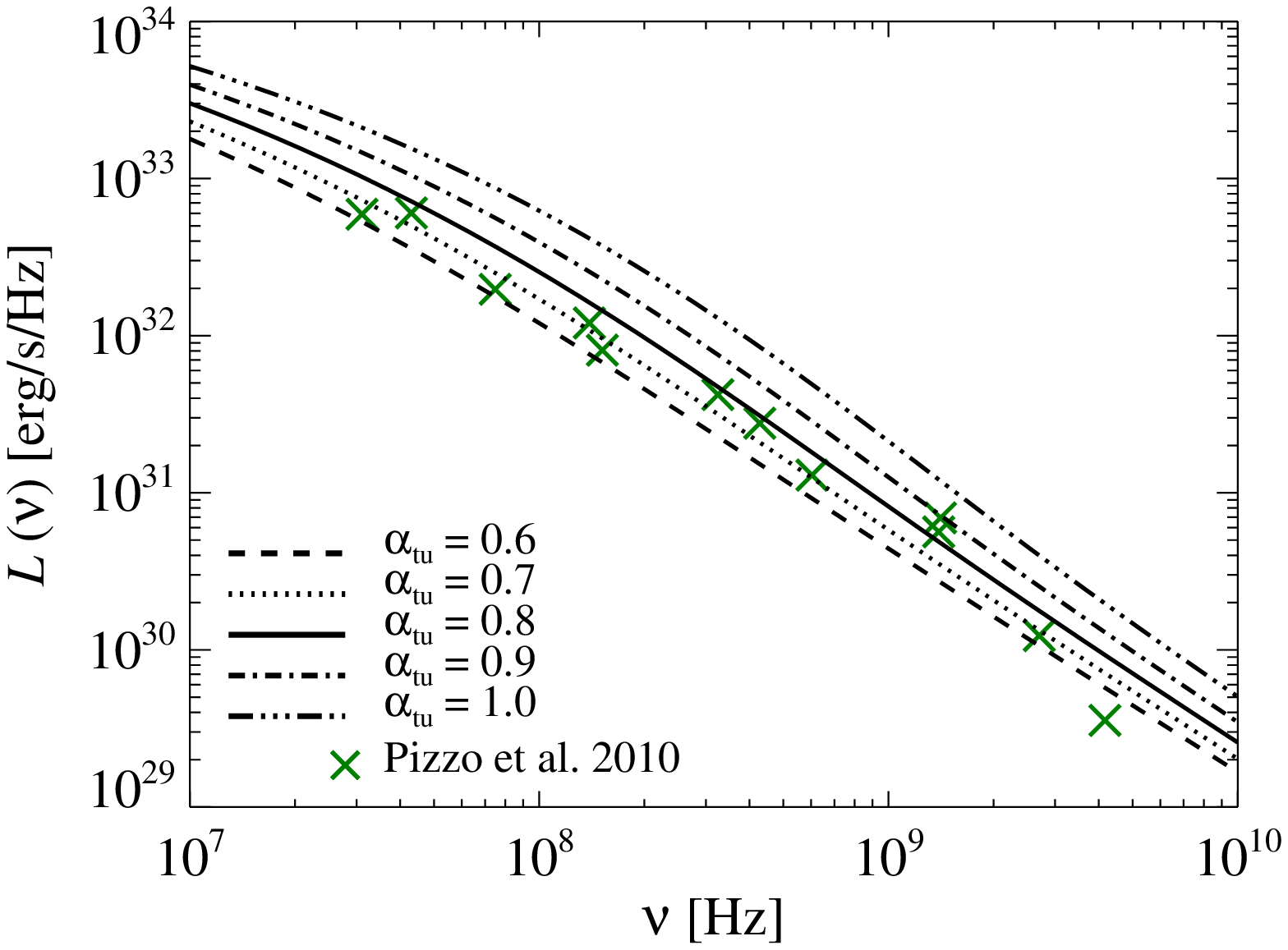}
   \end{center}
\end{minipage}
\\
\begin{minipage}{1\columnwidth}
  \begin{center}
    \includegraphics[width=\columnwidth]{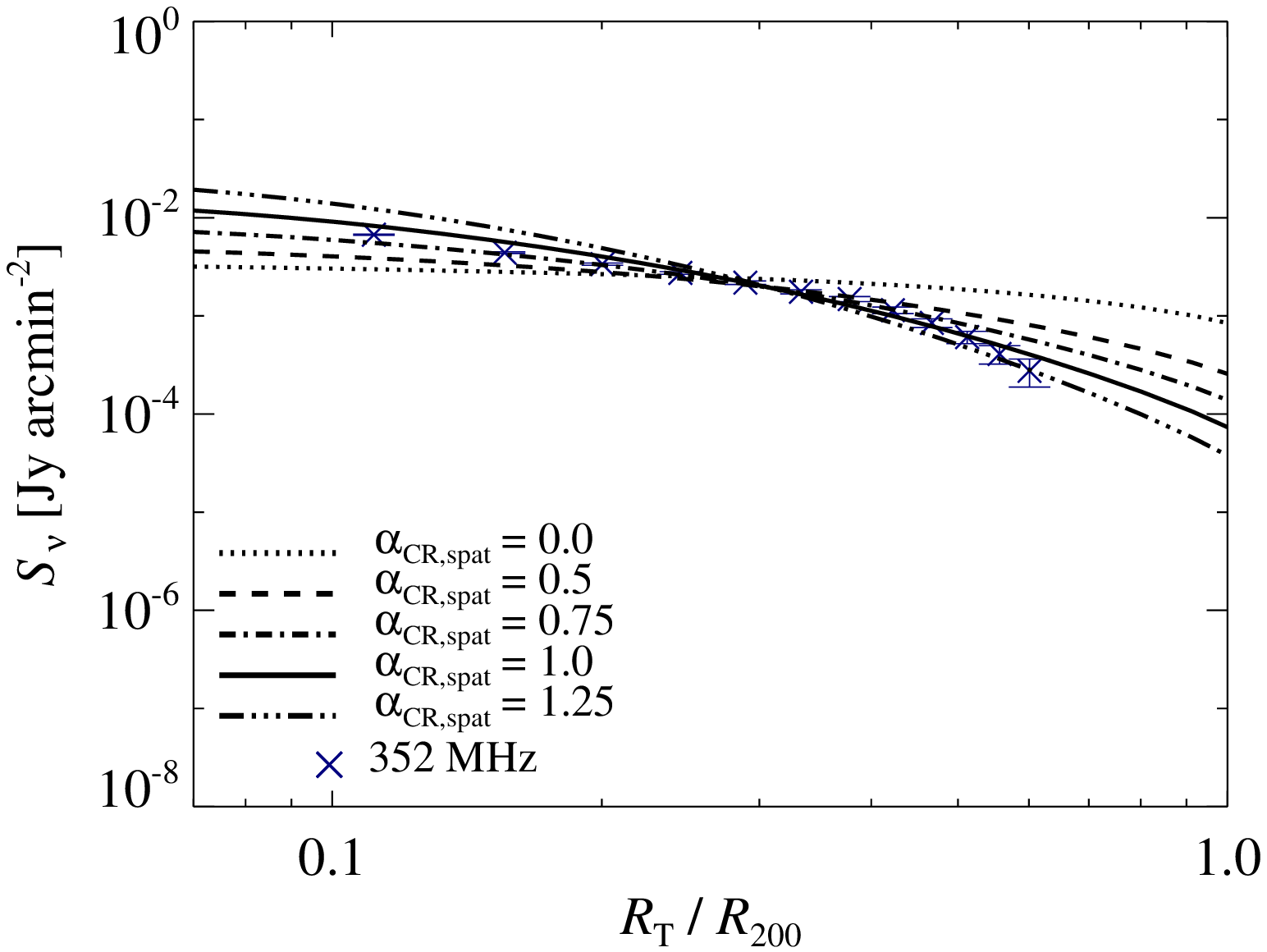}
  \end{center}
\end{minipage}
\begin{minipage}{1\columnwidth}
   \begin{center}
     \includegraphics[width=\columnwidth]{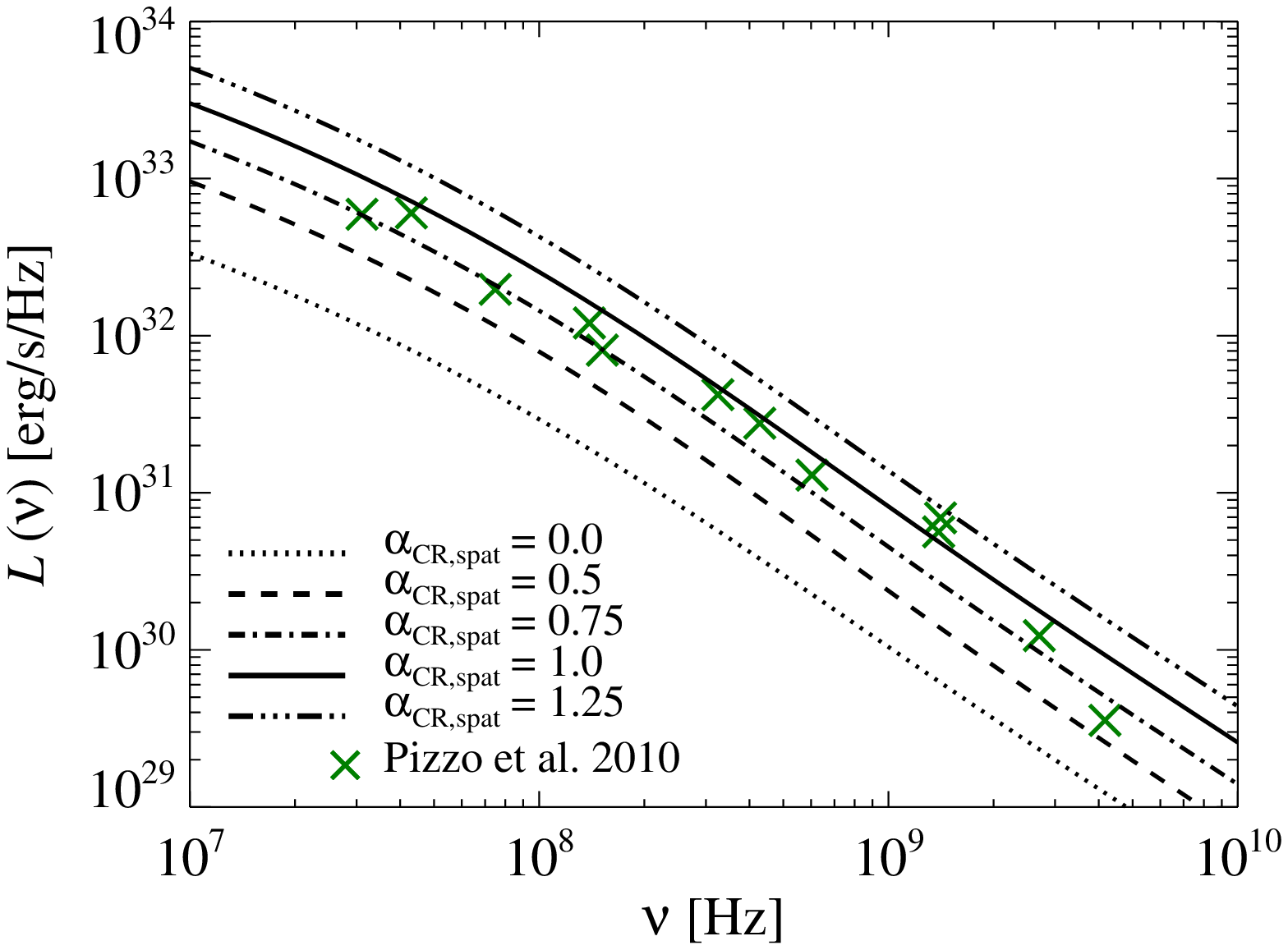}
   \end{center}
\end{minipage}
\caption{Sensitivity of the radio emission in the Coma cluster to
  critical parameters. The left-hand panels show the radio surface
  brightness profiles. We compare profiles at 352~MHz \citep[blue
    crosses,][]{brown11} to predicted emission from Fermi-II
  reaccelerated CR electrons (black lines). The right-hand panels show
  radio synchrotron spectra. The green crosses are compiled from
  observations \citep{2010PhDT.......259P}, while the black lines show
  predicted spectra. The upper panels show the sensitivity to the
  level of turbulence ($X_\rmn{tu}$), the middle panels show the
  impact of different turbulent profiles ($\alpha_\rmn{tu}$), and the
  lower panels show the dependence on spatial distributions of initial
  and injected CRs ($\alpha_\rmn{CR,spat}$). We adopt the following
  fiducial values for our model (solid lines), $X_\rmn{tu}=0.2$,
  $\alpha_\rmn{tu}=0.8$, and $\alpha_\rmn{CR,spat}=1.0$ and vary each
  parameter separately in each row of panels. We find that the radio
  emission is most sensitive to the level of turbulence. The abundance
  of CR seeds, and the spatial distribution of CRs and turbulence are
  second-order effects.}
  \label{fig:param_comp}
\end{figure*}

Figure~\ref{fig:param_comp} shows the impact of turbulence and the
spatial distribution of CRs on the radio emission. 

{\bf Impact of overall level of turbulence ($X_{\rm tu}$).} From the top panels of Figure~\ref{fig:param_comp}, we see that as $X_{\rm tu}$ increases, there are 3 important changes: an exponential increase in radio luminosity, a flattening of the radio surface brightness profile, and an increase in spectral curvature. We discuss these in turn. 

As we show in Section~\ref{sect:self-limiting}, the exponential increase in radio surface brightness is easily understood from equation~\ref{eq:Creacc}, since for a power-law initial distribution function $f_{\rm i}(p) = C_{\rm 0} p^{-\alpha_{\rmn{inj}}}$, then $C_{\rm reacc} \propto C_{\rm 0} {\rm exp}(A \tau_{\rm cl}/\tau_{\rm D}) \propto {\rm exp} (B(r) X_{\rm tu}^{2})$. This exponential sensitivity is somewhat modified by cooling (which results in a non power-law spectrum; also, the shorter the acceleration time, the larger the pool of seed electrons available which would otherwise cool away), but overall is a good approximation.  

An increase in $X_{\rm tu}$ flattens the surface brightness profile, since high acceleration efficiency leads to larger amplification in the cluster outskirts (where cooling is less important) than the centre. In particular, in the cluster outskirts, the reduced impact of Coulomb cooling implies that there is a larger pool of low-energy electrons available for reacceleration (see timescales in Table \ref{tab:timescales}). 

Larger levels of turbulence also {\it increase} spectral curvature, which might seem puzzling. It can be understood as follows. The pre-acceleration electron distribution function results from the competition between hadronic injection $f \propto p^{-\alpha_{\rm inj}}$ and cooling, which results in a quasi-steady state for the non-thermal (secondary) electrons. At low momenta, when the Coulomb cooling time is short, $f \propto p^{-\alpha_{\rm inj}+1}$, at high momenta, when inverse Compton/synchrotron cooling dominates, $f \propto p^{-\alpha_{\rm inj} -1}$. In between, there is a quasi-adiabatic regime where electrons accumulate (for more details, and an analytic self-similar solution, see \citealt{1999ApJ...520..529S, pinzke13}). In the absence of cooling, momentum advection in the limit $D_{pp} \propto p^{2}$ (so that $\tau_\rmn{D} =p^{2}/4 D_{pp}$ is momentum independent) simply shifts the distribution $f(p) \rightarrow f(A p)$. 
When the acceleration efficiency is low, most observable emission corresponds to the power-law tail of the distribution function $f \propto p^{-\alpha_{\rm inj}-1}$ set by the balance between injection and synchrotron/IC cooling. However, as the acceleration efficiency $A$ increases, radio emission starts to probe the 'bump' around $p_{*}$ (given by $\tau_{\rm D} \sim \tau_{\rm cool}(p_{*})$) where electrons accumulate and the distribution function is curved. This results in a curved emission spectrum. The synchrotron spectrum steepens at the frequency \citep{2001MNRAS.320..365B}: 
\begin{equation}
\nu_{\rm s} \propto \frac{B \tau_{\rm D}^{-2}}{(B^{2}+B^{2}_{\rm CMB})^{2}}
\end{equation}
where $B_{\rm CMB} \equiv (8 \upi \eps_{\rm CMB})^{1/2}$, which increases for shorter $\tau_{\rm D}$.  

{\bf Impact of spatial profile of turbulence ($\alpha_{\rm tu}$).} From the middle panels of Figure~\ref{fig:param_comp}, we see that as expected, a flatter profile of the turbulent pressure directly translates into a flatter radio surface brightness profile. Since seed electrons are more concentrated toward the centre (the collisional production of secondaries is more rapid there), and magnetic fields are stronger, concentrating the turbulence toward the cluster centre for fixed $X_{\rm tu}$ results in higher radio luminosities, and slightly more curvature (due to the increased importance of cooling near the centre). Overall, however, the spatial profile of turbulence has a much weaker effect than its overall normalisation. 

{\bf Impact of spatial profile of seed CRs $\alpha_{\rm CR, spat}$.} The spatial distribution of CRs has an even smaller impact on radio emission. At fixed total CR energy content $X_{\rm CR}$, concentrating the CRs towards the centre leads to more centrally dominated surface brightness profiles, as expected, and higher radio luminosities (for the same reasons as above: secondaries are more easily produced in the centre, and magnetic fields are stronger). We have also confirmed that radio surface brightness profiles scale linearly with $X_{\rm CR}$, as expected. 

Overall, our results suggest that radio haloes are much more sensitive to the level of turbulence (exponential dependence) rather than CR abundance (linear dependence), and that the spatial distribution of turbulence and CRs, while important, are second-order effects. In our parametrization, the most important controlling variable is $X_{\rm tu}$. The overall level of turbulence has to be such that $\tau_{\rm D} \sim \tau_{\rm cl}$ (see Table \ref{tab:timescales}), otherwise, too little or too much amplification takes place. For instance, for $X_\rmn{tu}\gtrsim0.08$, changing $X_\rmn{tu}$ by a factor of two changes the radio surface brightness by a factor of $\sim 10-100$ (see top panels of Figure~\ref{fig:param_comp}). The required $\tau_{\rm D}/\tau_{\rm cl}$ depends only logarithmically on the abundance of seed CRs.

While the requirement of a threshold level of turbulence may explain why radio brightness is bimodal, it also raises a fine-tuning problem: why is the $L_{\rm radio}$ vs. $L_{\rm X}$ relation in active radio haloes so tight \citep{Brunetti2009}? Depending on the details of infall or mergers, we would naturally expect fluctuations in $X_{\rm tu}$, which would translate into large scatter in the $L_{\rm radio}$--$L_{\rm X}$ relation. This can be only be understood if the timescale over which acceleration takes place $\tau_{\rm cl}$ also depends on the properties of turbulence, so that the ratio $\tau_{\rm D}/\tau_{\rm cl}$ has relatively little scatter. We address this next. 

\subsection{Self-limiting turbulent reacceleration}
\label{sect:self-limiting} 

In this section, we explore the physical origin of the high sensitivity of radio emission to turbulence levels (e.g., top right panel of Fig. \ref{fig:param_comp}), which thus requires strong fine-tuning to explain observed radio profiles. We will find that by relating the acceleration time to the turbulent decay time, this sensitivity can be eliminated.  We also challenge the common assumption of assuming a Kraichnan spectrum, beginning at the outer scale (see Section~\ref{sec:reacc}). Since turbulence is essentially hydrodynamic at large scales, this is unjustified. Instead, we shall assume that the outer scale of the compressive fast modes is the Alfven scale.

There are 3 important timescales in this problem: the acceleration time $\tau_{\rm D}$, the duration over which turbulence is active and acceleration takes place, $\tau_{\rm cl}$, and the cooling time $\tau_{\rm cool}(p)$. Only $\tau_{\rm cool}(p)$ is momentum dependent (and is different for ions and electrons). Thus, the outcome of acceleration depends essentially on two dimensionless numbers, $\tau_{\rm cl}/\tau_{\rm D}$, and $\tau_{\rm cool}/\tau_{\rm D}$. 

We have seen that the radio luminosity depends very sensitively on $\tau_{\rm
  cl}/\tau_{\rm D}$, through the very sensitive dependence on $X_{\rm tu}$ (Figure
\ref{fig:param_comp}; note from equation (\ref{eq:Dpp_scaling}) that $\tau_{\rm D}
\propto X_{\rm tu}^{-2}$). Since this raises questions of fine-tuning in $X_{\rm
  tu}$ to explain the observations, it is worth understanding this property in
more detail. To this end, we ignore cooling, which is a good approximation for
the CR protons, because hadronic cooling times are long in comparison to the
other relevant timescales of the problem.\footnote{In principle, CRp can still
  lose energy via wave heating, at a rate $\left|\bmath{\varv_{\rm A} \cdot
    \nabla} P_\CR\right|$, but we assume CR streaming is suppressed during
  mergers, when they are spatially confined by scattering.} In this case,
$\dot{p}= p/\tau_{\rm D}$, and after a time $\tau_{\rm cl}$, we have
\begin{equation}
  \label{eq:pexp}
p\rightarrow p \, {\rm exp}(\tau_{\rm cl}/\tau_{\rm D})
\end{equation}
(where we have used the fact that $\tau_{\rm D}$ is momentum independent). For
an initial power law distribution function $f(p) = C p^{-\alpha_{\rmn{inj}}}$,
this momentum increase can be rewritten as a change of normalisation, $f(p) =
\tilde{C} p^{-\alpha_{\rmn{inj}}}$, where $\tilde{C} = {\rm exp}(
\alpha_{\rmn{inj}} \tau_{\rm cl}/\tau_{\rm D})$. A slightly more careful
derivation by direct solution of the Fokker-Planck equation yields:
\begin{equation}
  \dot{f} = {{\partial }\over{\partial p}}
  p^2 D_{pp} {{\partial }\over{\partial p}} \frac{f}{p^2}\,,
\end{equation}
with the analytic solution given by

\begin{equation}
  \label{eq:Creacc}
  \tilde{C}= 
  C \exp{\left[\frac{(2+\alpha_{\rmn{inj}})(\alpha_{\rmn{inj}}-1)}{4}\frac{t}{\tau_\rmn{D}}\right]}\,.
\end{equation}
At $t=\tau_\rmn{cl}$ the CR distribution is exponentially sensitive to
$\tau_{\rm cl}/\tau_{\rm D}$, which is the physical reason underlying
the extreme sensitivity to $X_{\rm tu}$ in
Figure~\ref{fig:param_comp}.

The natural solution to this puzzle would be a process that couples
$\tau_{\rm cl}$ and $\tau_{\rm D}$. There are two possible limits. Let
us suppose that the timescale on which there exists a source of
turbulent driving is $\tau_{\rm drive}$, which approximately
corresponds to the timescale of the merger, or the dynamical time. Let
the turbulence decay on a timescale $\tau_{\rm decay}$. If $\tau_{\rm
  drive} > \tau_{\rm decay}$, then $\tau_{\rm cl} \sim \tau_{\rm
  drive}$, which depends on the details of the merger and should
result in considerable scatter in $\tau_{\rm cl}/\tau_{\rm D}$ in
different systems. On the other hand, if $\tau_{\rm decay} > \tau_{\rm
  drive}$, then $\tau_{\rm cl} \sim \tau_{\rm decay}$. Since
$\tau_{\rm D}$ and $\tau_{\rm decay}$ are both related to properties
of the turbulence, it is conceivable that this would result in much
less scatter in $\tau_{\rm cl}/\tau_{\rm D}$.

An obvious candidate for $\tau_{\rm decay}$ is: 
\begin{equation}
\tau_{\rm edd} = \frac{L}{\varv_{\rm L}}
\end{equation}
i.e., the eddy turnover time at the outer scale. This is subject to
uncertainties about the location of the outer scale $L$; estimates in the
literature range from $L \sim 0.1-1$ Mpc. It is also worth remembering that MHD
turbulence only applies for $l < l_{\rm A}$, where $l_{\rm A}$ is the Alfv{\'e}n
scale where $\varv \sim \varv_{\rm A}$. Invoking fast modes, Kraichnan scalings,
etc., is only valid below these scales. For $l > l_{\rm A}$, turbulence is
basically hydrodynamic.  

In the hydrodynamic regime, a standard Hodge-Helmholtz decomposition usually shows that the compressive component of the velocity field is Burgers-like ($W(k) \propto k^{-2}$), while the solenoidal component is Kolmogorov-like ($W(k) \propto k^{-5/3}$), see e.g., \citet{federrath13}. The Burgers-like component does not reflect a genuine cascade, but rather the appearance of shocks which directly transfer power from large to small scales. At face value, we should use the Burgers spectrum for compressible modes. However, as already found by previous authors \citep{miniati15, brunetti16}, and as we shall discuss, this does not produce significant particle acceleration. If this is the correct spectrum, then the paradigm of turbulent particle acceleration is simply flawed, and some other mechanism is necessary to explain radio halos. Alternatively, it is well-known that due to mode-mode coupling,  solenoidal modes can give rise to compressive modes, and vice-versa \citep{kida90,cho03,kritsuk07}, even for subsonic turbulence, since pressure fluctuations of order $\sim \rho u^{2}$ arise. Indeed, note that many numerical studies (such as \citet{cho03,kritsuk07}) only have solenoidal driving on the outer scale, but then are also able to study the compressive modes that develop. In hydrodynamic turbulence, the energy in compressible modes which develop in this way scales as $\sim \mathcal{M}_{s}^2$, for $\mathcal{M}_{s} < 1$; this coupling is strongest at the Alfven scale \citep{cho03}. In MHD turbulence, mode-mode coupling is weak below the Alfven scale and the Alfven, fast and slow modes proceed as separate cascades.

We thus assume that some fraction of the Kolmogorov-like hydrodynamic turbulence which cascades down to the Alfven scale ends up as compressive fast modes with a Kraichnan spectrum, as seen in simulations \citep{cho03}. Henceforth, we can consider the outer scale of the fast modes to be $l_{\rm A}$. If some fraction $f_{\rm c}$ of the turbulent energy density at this scale is in compressible modes, then the energy density at the outer scale is $\sim f_{\rm c} \rho \varv_{\rm A}^{2}$.
The turbulent reacceleration time is:

\begin{equation}
\tau_{\rm D} = \frac{p^{2}}{4 D_{pp}} = \frac{C_{\rm D}}{A^{1/2}}\, \frac{c}{k_{\rm A}}\, \frac{\beta}{f_{\rm c} \varv_{\rm A}^{2}}
\label{eqn:tau_D} 
\end{equation}

where $C_{\rm D} = 2/(5 \upi)$, $A\approx 11000$, and we have used equations, (\ref{eqn:diffusion}), (\ref{eqn:k_W}), (\ref{eqn:k_c}), assuming $s=3/2$. The turbulent decay time is given by the cascade time of fast modes at the Alfv\'en scale $k=k_{\rmn{A}}$ (\citealt{2004ApJ...614..757Y}; see equation \ref{eqn:tau_l}): 
\begin{equation}
  \tau_{\rm decay} = \frac{\varv_{\rm ph}}{\varv_k^{2} k} = 
\frac{c_{\rm s}}{f_{\rm c} \varv_{\rm A}^{2} k_{\rm A}^{}}\,.
\label{eqn:tau_decay} 
\end{equation}
This can be related to the eddy turnover time at the outer scale $\tau_{\rm edd}$ from $l_{\rm A} = L \mathcal{M}_{\rm A}^{-3}$ to: 
\begin{equation}
\tau_{\rm decay} =\frac{L}{\varv_{\rm L}} \frac{c_{\rm s}^{2}}{f_{\rm c} \varv_{\rm L}^{2}}  \frac{\varv_{\rm A}}{c_{\rm s}} \approx  1.0 \, \tau_{\rm edd} \left(\frac{\tilde{X}_{\rm tu}}{0.2}\right)^{-1}  \left(\frac{\beta}{50} \right)^{-1/2}
\end{equation}
where we have used $\tilde{X}_{\rm tu} \approx f_{\rm c} (\varv_{\rm L}/c_{\rm s})^{2}$ and $c_{\rm s}/\varv_{\rm A} = \beta^{1/2}$. The decay time at the Alfven scale is comparable (and could be larger, if $f_{\rm c}$ is smaller than we have assumed) to the eddy turnover time at the outer scale. Thus, it is appropriate to consider the latter as the decay time for the fast modes, rather than the cascade time at the driving scale. The reason for these comparable timescales despite the disparate length scales is that large scale hydrodynamic modes cascade on a single eddy turnover time, whereas small scale sub-Alfvenic MHD modes are in the weak turbulence limit, and require multiple wave-wave interactions for a mode to cascade.  

From equations (\ref{eqn:tau_D}) and (\ref{eqn:tau_decay}), we obtain:
\begin{equation}
  \frac{\tau_{\rm cl}}{\tau_{\rm D}} \approx \frac{\tau_{\rm decay}}{\tau_{\rm D}}
  = \frac{A^{1/2}}{C_{\rmn{D}}}\,\frac{c_{\rm s}}{c}\,\beta^{-1}
  = 0.1 \left( \frac{c_{\rm s}}{1500 \, {\rm km \, s^{-1}}} \right) \left( \frac{\beta}{50} \right)^{-1}\,. 
\label{eqn:ratios_TTD} 
\end{equation}
Remarkably, this expression is independent of properties of the turbulence such as $\varv_{\rm c}$, $L$, or $l_{\rm A}$, and depends only on properties of the plasma ($c_{\rm s},\beta$). Ultimately, this arises because the timescale on which a fast mode wave transfers energy due to wave particle interactions in second order Fermi acceleration, $\tau_{\rm p} \sim [4\, k_{\rm p} c (\varv_{\rm c}/c)^{2} ]^{-1}$, is closely related to the timescale on which it cascades due to wave-wave interactions, $\tau_{\rm w} \sim c_{\rm s}/(k_{\rm w} \varv_{\rm c}^{2})$. This implies $\tau_{\rm w}/\tau_{\rm p} \sim 4\, (c_{\rm s}/c)(l_{\rm w}/l_{\rm p})$, where $l_{\rm w} \sim l_{\rm A}$ is the outer scale on which the fast mode cascade begins, and $l_{\rm p} \sim (l_{\rm A} l_{\rm cut})^{1/2}$ is the characteristic wavelength for wave-particle interactions. For transit time damping and an outer scale of $l_{\rm A}$, we have $l_{\rm cut} \sim (m_{\rm e}/m_{\rm p}) l_{\rm A} (c_{\rm s}/\varv_{\rm A})^{4}$. We thus have $\tau_{\rm w}/\tau_{\rm p} \sim 4\,(c_{\rm s}/c)(l_{\rm A}/l_{\rm cut})^{1/2} \sim 4\,(c_{\rm s}/c) (m_{\rm p}/m_{\rm e})^{1/2} \beta^{-1}$. More careful consideration of dimensionless factors boosts this estimate by an order of magnitude to give equation (\ref{eqn:ratios_TTD}). 

On the other hand, equation (\ref{eqn:ratios_TTD}) points toward a pessimistic scenario where turbulent reacceleration with TTD on thermal particles is never effective. A key reason is that we assume Kraichnan turbulence only applies for $l< l_{\rm A}$. There is then insufficient separation of scales: the cutoff scale $l_{\rm cut} \sim 0.2 \beta_{\rm 50}^{2} l_{\rm A}$. Although there is a fair large separation of scales between the outer driving scale $L$ and $l_{\rm A} = L \mathcal{M}_{\rm A}^{-3}$ (a factor of $30-1000$ for $\mathcal{M}_{\rm A} \approx 3-10$; we have $\mathcal{M}_{\rm A} \sim (\tilde{X}_{\rm tu} \beta)^{1/2} \sim 3.2$ for our fiducial assumptions), Kolmogorov turbulence, with its steeper spectrum, has more energy at large scales ($k W(k) \propto k^{-2/3}, k^{-1/2}$ for Kolmogorov and Kraichnan turbulence, respectively). This implies that the energy-weighted scales at which wave-particle interaction take place are larger in Kolmogorov turbulence, and thus that the wave-particle interaction rate is lower. If (as is frequently seen) we instead assume that Kraichnan turbulence begins at the outer scale $L$, with characteristic decay time $L c_{\rm s}/\varv_{\rm c}^{2}$, then we obtain a more palatable result: 
\begin{equation}
\frac{\tau_{\rm cl}}{\tau_{\rm D}} \approx 0.8  \left( \frac{c_{\rm s}}{1500 \, {\rm km \, s^{-1}}} \right) \left( \frac{\tilde{X}_{\rm tu}}{0.2} \right). 
\end{equation}
This arises because the decay time is now $\sim \mathcal{M}_{\rm A}$ times longer, and the acceleration time is now $\sim \mathcal{M}_{\rm A}^{1/2}$ times shorter, boosting $\tau_{\rm cl}/\tau_{\rm D}$ by $\sim \mathcal{M}_{\rm A}^{3/2}$. However, as we have argued, turbulence in the hydrodynamic regime $l_{\rm A} < l < L$ is Kolmogorov, not Kraichnan. Furthermore, if this scaling is correct, then this leaves us with the problematic exponential sensitivity to $X_{\rm tu}$ that we previously explored. 

One alternative is that scattering in the high-$\beta$ ICM is mediated by plasma instabilities (firehose, mirror) rather than Coulomb scattering. This vastly increases the scattering rate and reduces the mean free path of thermal particles. In this case, the fast modes damp by TTD on relativistic rather than thermal particles. The momentum diffusion coefficient in this case is then \citep{brunetti11, miniati15}:
\begin{equation}
D_{pp}^{\rm CR} = \frac{2 p^{2} \zeta}{X_{\rm CR}} k_L \frac{\langle \varv_{\rm c}^{2} \rangle^{2}}{c_{\rm s}^{3}}
\end{equation}
where $\zeta$ is an efficiency factor for the effectiveness of plasma instabilities (e.g., due to spatial or temporal intermittency), and $X_{\rm CR}=\eps_{\rm CR}/\eps_{\rm th}$ is the relative energy density of CRs. If we set $k_L \rightarrow k_{\rm A}$, $\varv_{c}^{2} \rightarrow \varv_{\rm A}^{2}$, as before, then: 
\begin{equation}
\frac{\tau_{\rm cl}}{\tau_{\rm D}} \approx 8 \frac{\zeta}{\beta X_{\rm CR}}
\end{equation}
In principle, this implies exponential sensitivity to $X_{\rm CR}$, a similar situation as exponential sensitivity to $X_{\rm tu}$. However, there is a self-limiting asymptotic behaviour: as $X_{\rm CR}$ increases due to turbulent reacceleration, damping increases, which limits the further growth of the CR energy density \citep{brunetti11}. One natural assumption is to assume that $\eps_{\rm CR}$ saturates at a level $\eps_{\rm CR} \sim \eps_{\rm turb} ({\rm MHD}) \sim \eps_{\rm B}$, in which case $X_{\rm CR} \sim \beta^{-1}$, so that: 
\begin{equation}
\frac{\tau_{\rm cl}}{\tau_{\rm D}} \approx 2.5 \left(\frac{\zeta}{0.3}\right)\,.
\label{eqn:ratios_CR_damp} 
\end{equation}
Given the host of uncertain factors which enter into $\zeta < 1$ (the intermittent nature of turbulence, small scale magnetic topology and the efficiency with which instabilities mediated by pressure anisotropies are triggered), this is potentially consistent with $\tau_{\rm cl} \sim \tau_{\rm D}$. This promising scenario should be investigated in more detail. 

It is also worth exploring the effect of Burgers' turbulence (weak shocks). Burgers' turbulence is distinct from other forms of turbulence in that it violates locality in $k$-space--i.e., power does not gradually cascade, but instead can directly jump from large to small scales via a network of weak shocks. This implies a short dissipation time and we have in this case, $\tau_{\rm cl} \sim \tau_{\rm drive}$, the driving time, rather than the turbulent dissipation time. Since the Fourier transform of a step function goes as $k^{-1}$, Burgers turbulence has a power spectrum of $P(k) \propto k^{-2}$. From equations (\ref{eqn:diffusion}) and (\ref{eqn:k_W}), we find: 
\begin{equation}
\frac{D_{pp}^{\rm Burgers}}{D_{pp}^{\rm MHD}} = \left( \frac{\beta}{X_{\rm tu} A} \right)^{1/2} = 0.15 \left( \frac{\beta}{50} \right)^{1/2} \left( \frac{X_{\rm tu}}{0.2} \right)^{-1/2} 
\end{equation}
where we have set $\langle k \rangle_{\mathcal{W}} \approx k_L$ for Burgers' turbulence, i.e. all power is at large scales. This weighting towards larger scales implies a much lower wave-particle interaction rate and thus a diffusion coefficient which is smaller than standard TTD on thermal particles by an order of magnitude. Since $\langle k \rangle_{\mathcal{W}} \approx k_L$ is independent of the cutoff scale $k_{\rm cut}$, this conclusion is unchanged if plasma instabilities regulate the thermal particle mean free path and TTD operates on relativistic particles instead. Since standard TTD on thermal particles was already potentially problematic (equation \ref{eqn:ratios_TTD}), we conclude, in agreement with other assessments \citep{miniati15,brunetti16_review}, that if Burgers turbulence dominates, then turbulent reacceleration will be ineffective.

\begin{figure*}
\begin{minipage}{1\columnwidth}
   \begin{center}\Large{radio profiles}\\
     \includegraphics[width=\columnwidth]{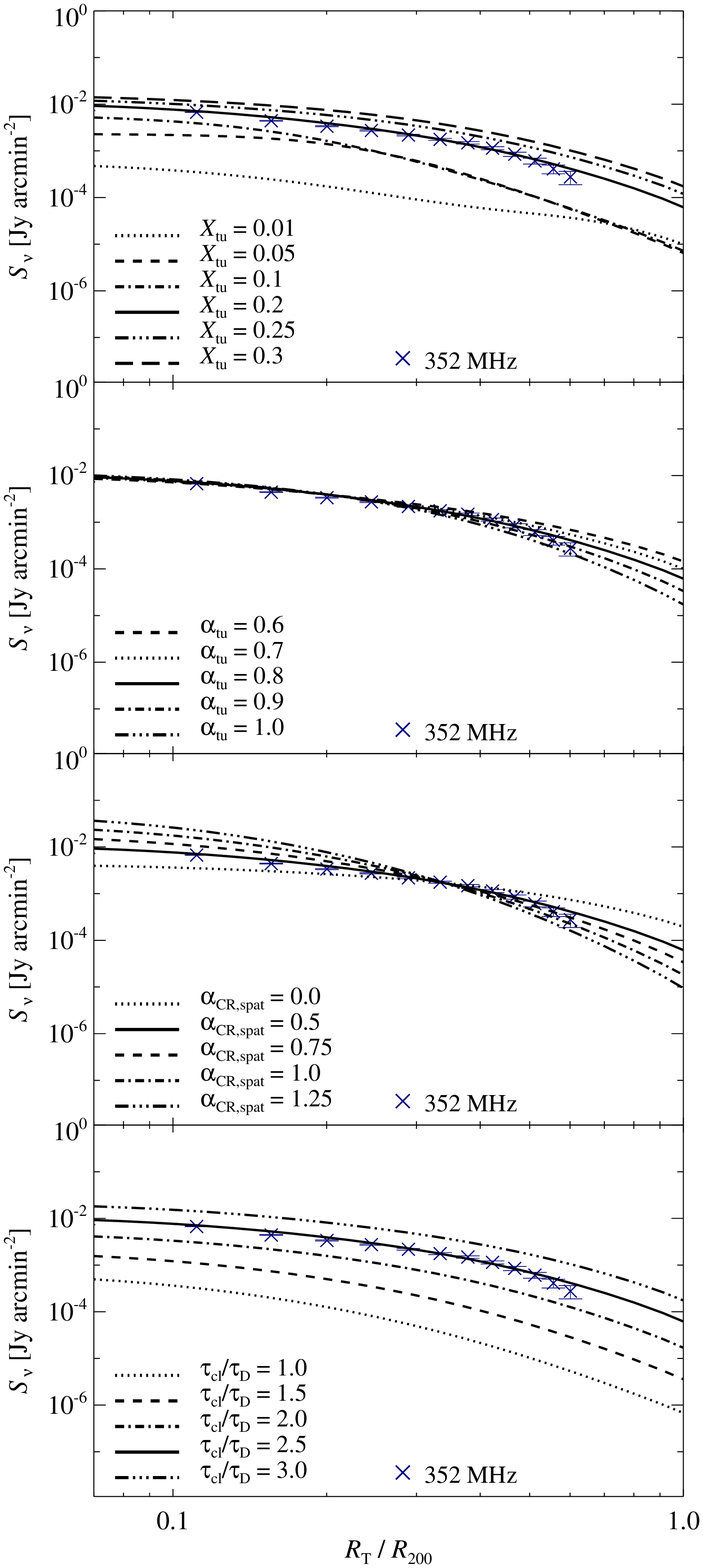}
   \end{center}
\end{minipage}
\begin{minipage}{1\columnwidth}
   \begin{center}\Large{radio spectra}\\
     \includegraphics[width=\columnwidth]{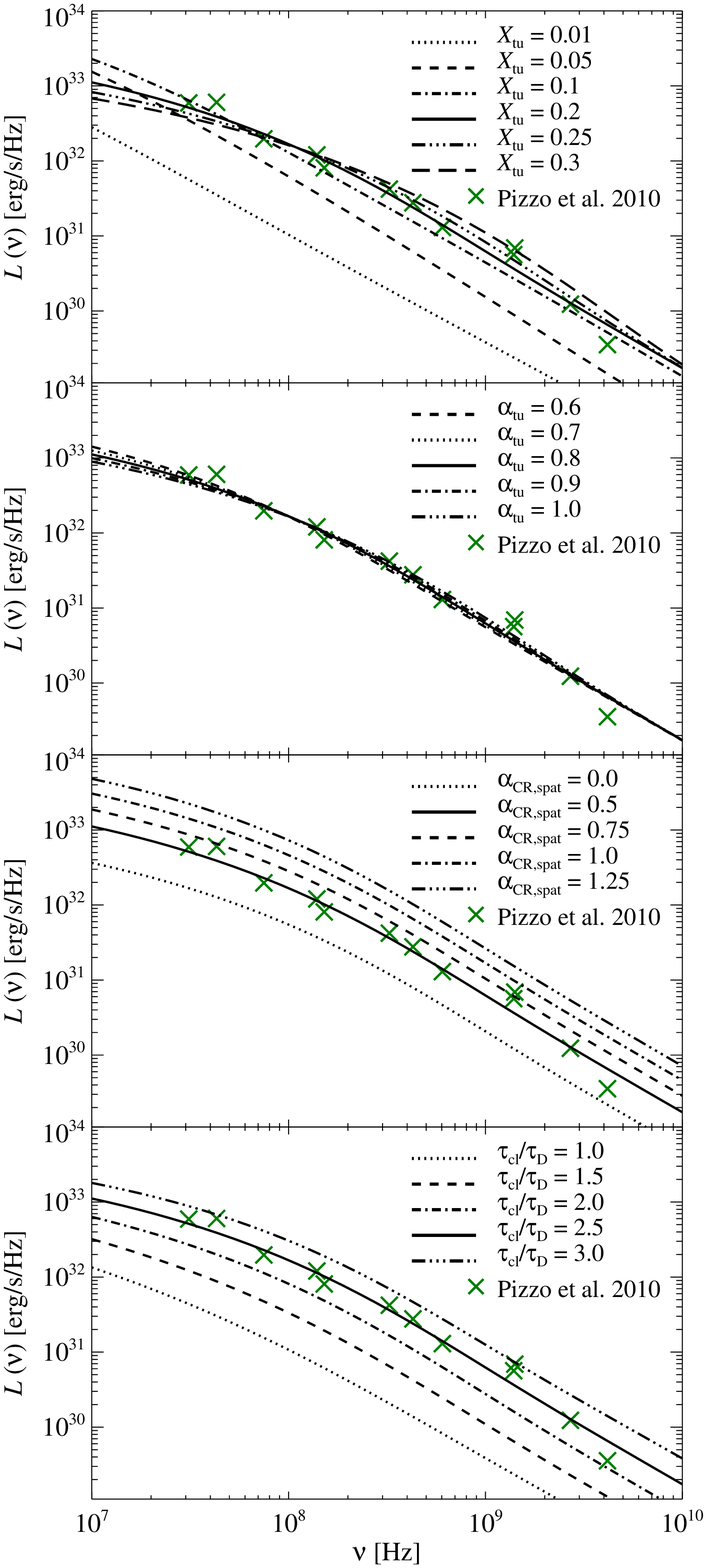}
   \end{center}
\end{minipage}
\caption{Sensitivity of radio emission in the Coma cluster to critical
    parameters when we vary the ratio of the time scales on which turbulence is
    active ($\tau_{\rm cl}$) to the turbulent re-acceleration time ($\tau_{\rm
      D}$). In comparison to Figure~\ref{fig:param_comp}, we change our fiducial
    model (solid lines) and adopt a slightly more extended spatial distribution
    of CRs ($\alpha_{\rm CR, spat}=0.5$) and lower the injection rate of CR
    energy by a factor five; otherwise, we adopt $X_\rmn{tu}=0.2$,
    $\alpha_\rmn{tu}=0.8$, $\tau_{\rm cl}=2.5\,\tau_{\rm D}$ as fiducial values
    and vary each parameter separately in each row of panels. Radio surface
    brightness profiles (left) are contrasted to radio synchrotron spectra
    (right; for more details see caption to Figure~\ref{fig:param_comp}).
    Provided there is a minimum (threshold) level of turbulence
    ($X_\rmn{tu}\gtrsim0.2$), here the radio profiles are mainly determined by
    the CR distribution, while turbulence has a remarkably small impact. In the
    bottom two panels we vary the exponential momentum growth factor $\tau_{\rm
      cl}/\tau_{\rm D}$ (see equations~\ref{eq:pexp} and \ref{eq:Creacc}), which
    mainly changes the normalisation of the CR distribution and neither spectral
    shape nor radial synchrotron profile. Interestingly, the flat CR profiles
    adopted here are similar to those required to reproduce the radio profiles
    in our CR streaming model (see Figure~\ref{fig:sync_profile}).}
  \label{fig:param_comp_tcl_tD}
\end{figure*}

It is interesting to reconsider surface brightness and spectral profiles if
  indeed $\tau_{\rm D} \sim \tau_{\rm cl}$ for the reasons mentioned above
  (e.g., equation~\ref{eqn:ratios_CR_damp}). We show the results of adopting
  such an ansatz in Figure~\ref{fig:param_comp_tcl_tD}, where, similar to
  Figure~\ref{fig:param_comp}, we vary $X_{\rm tu}$, $\alpha_{\rm tu}$, and
  $\alpha_{\rm CR, spat}$ about fiducial values.\footnote{When $\tau_{\rm
        cl}/\tau_{\rm decay} \sim\rmn{const}$ is adopted, the radial behaviour
      of $\tau_{\rm cl}$ is also modified. Because $\tau_{\rm D}$ decreases with
      radius (see e.g. Figure~\ref{fig:tauD}), $\tau_{\rm cl}$ has to decrease
      by the same amount, which results in a steeper radio profile. To
      compensate for this effect in Figure~\ref{fig:param_comp_tcl_tD}, we
      change $\alpha_{\rm CR, spat}=1.0\rightarrow0.5$ and decrease the
      injection rate of CR energy by a factor five.} Instead of adopting a
  fixed $\tau_{\rm cl}$, we set $\tau_{\rm cl}/\tau_{\rm D} = 2.5 $. The results
  are remarkably revealing. In this case, we are indeed less sensitive to the
  properties of the turbulence, and more sensitive to the details of the CR seed
  population.  This gives hope to the possibility that one could effectively
  marginalise over the very uncertain properties of turbulence at larger scales
  in merging clusters (which are unlikely to be more precisely constrained
  observationally in the near future) to learn something about the underlying CR
  population.

For instance, in the third row of panels of
Figure~\ref{fig:param_comp_tcl_tD}, we see that once turbulence exceeds a
threshold value $X_{\rm tu} \sim 0.2$, then its overall exact energy density
does not matter -- profiles simply converge to an asymptotic form as $X_{\rm
  tu}$ is increased. This is in contrast to the exponential sensitivity,
flattening in surface brightness, and increasing spectral curvature as $X_{\rm
  tu}$ is increased that we saw previously. A threshold value of $X_{\rm tu}$ is
necessary for turbulent reacceleration to overcome cooling. Once that condition
is satisfied, the condition $\tau_{\rm cl}/\tau_{\rm D} \sim\rmn{const}$ implies a
fixed amount of amplification: stronger turbulence implies faster acceleration,
but also dissipates more quickly. This leads to the asymptotic behaviour
seen. The value of this constant depends on plasma physics details, but we
  have shown that the value required to match observations is plausible.

The second row of panels of Figure~\ref{fig:param_comp_tcl_tD} reveal another
  striking result: in contrast to Figure~\ref{fig:param_comp}, results are also
completely insensitive to the spatial profile of turbulence. Given the large
uncertainties, this is a welcome property. It arises because for our fiducial
value ($X_{\rm tu} = 0.2$), variations in the turbulent profile still lead to
local energy densities which are above the threshold value required to overcome
cooling, and turbulent amplification approaches its asymptotic value.

Finally, in the top panels of Figure~\ref{fig:param_comp_tcl_tD}, we see
that we {\it are} sensitive to the spatial profile of CRs; it affects both the
radio surface brightness profile (a flatter CR distribution implies flatter
surface brightness profiles) and luminosity (once again, luminosity is larger
for more centrally concentrated CRs, since more secondaries are produced). This
arises because once turbulent reacceleration overcomes cooling, the condition
$\tau_{\rm cl}/\tau_{\rm D} \sim\rmn{const}$ provides a fixed amount of amplification in
each radial shell. The radio luminosity in each shell then depends linearly on
initial conditions, i.e., the profile of CR seeds. Interestingly, the
best-fitting profile for Coma results from a flat CR distribution -- i.e., the
outcome of efficient CR streaming. The overall amplitude is set by $\tau_{\rm cl}/\tau_{\rm D}$ (equation
\ref{eqn:ratios_CR_damp}) and the CR content.

Obviously, the ideas in this section require further study. For instance, the notion that fast modes appear abruptly at $l_{\rm A}$ and that they last for a time $\tau_{\rm cl}$ are clear approximations (the latter could be improved by solving the time dependent diffusion equation for the fast mode energy spectrum, see e.g. \citet{zhou90}). However, the notion that $\tau_{\rm cl}/\tau_{\rm D}$ could self-regulate around a fixed value is exciting, because it eliminates the main source of uncertainty -- the poorly unconstrained properties of turbulence, and implies that radio haloes could potentially provide more robust constraints on the underlying CR population.  

\section{Conclusions}
\label{sec:conclusions}

The standard reacceleration model for radio haloes requires two ingredients with
no current direct observational constraints: a CR population which produces seed
electrons, and turbulence to perform second-order Fermi acceleration on these
electrons\footnote{Other model ingredients, such as temperature, density, and
  B-field profiles, are observationally constrained by X-ray and Faraday
  rotation measurements.}. For the best studied radio halo, Coma, there are two
main observational constraints: the radio surface brightness profile
$S_{\nu}(r)$, and the integrated luminosity as a function of frequency
$L(\nu)$. For certain assumptions about turbulence and the seed population,
analytic theoretical models can match these observations \citep{brunetti11}, but
the realism of the assumed seed electron and turbulent profiles must be
confronted with numerical simulations.

Cluster turbulence has been studied numerically in cosmological simulations with
an eye toward radio halo properties \citep{2013ApJ...771..131B,
  miniati15}, though they were not confronted against the two
benchmark observations mentioned above. Most importantly, direct simulation of the
seed electron population has been missing in the literature. In this paper, we
fill that gap. At the same time, we study the sensitivity of radio halo profiles
and spectra to variation in properties of turbulence and seed population, and
uncover the dominant effects.

In the first part of the paper, we use hydrodynamic zoom simulations of a
Coma-like cluster in a cosmological setting to generate spatially and
momentum-resolved seed populations of CRps and CRes from diffusive shock
acceleration. The simulations include time-dependent cooling and adiabatic
transport processes, and has been previously used in a variety of applications
in CR cluster physics \citep{pinzke10,pinzke13}. The resultant CRp and CRe
distribution functions then serve as initial conditions for our calculations of
Fermi II acceleration by turbulence. We use time-dependent Fokker-Planck
calculations which allow for spatial variations in the properties of turbulence
(i.e., a spatially varying momentum diffusion coefficient). We adopt a simple
parametric model for cluster turbulence, characterising it by its overall
normalisation, spatial profile, and characteristic lifetime $\tau_{\rm cl}$. Our
approach is relatively sophisticated in its treatment of the CR seed population
and relatively simplified in its treatment of the turbulence; it is thus
orthogonal and complementary to existing treatments which focus on the
time-dependent compressible turbulence \citep{miniati15}.

Given our seed electron population (which includes both primaries generated by DSA and secondary electrons created by CRps in hadronic interactions; for standard assumptions, the latter dominates), we find that applying the standard model previously used to reproduce Coma's radio halo properties \citep{brunetti11}, in which $\eps_{\rm turb} \propto \eps_{\rm th}$, fails: it produces radio emission which is too centrally concentrated. Indeed, \citet{brunetti11} remark in their paper that the CR population required in their model is remarkably flat. By contrast, the simulations produce CRs which are more centrally concentrated.

Thus, some modification of the standard model is needed, either by modifying the spatial distribution of the CRs, or that of turbulence. We explore two examples of the former: (i) CRs can stream in the ICM, producing a flat profile \citep{ensslin11,wiener13}, or (ii) given a higher e/p ratio in DSA acceleration (perhaps due to magnetic geometry; \citealt{2014ApJ...794..153G}), the primary population can dominate. This has a much flatter profile (since secondaries must be generated collisionally, they will always be concentrated towards denser regions). However, both of these solutions still require some modification of the turbulent profile, requiring $\eps_{\rm turb}/\eps_{\rm th}$ to rise with radius. Indeed, one can simply use the unmodified CR profile derived from simulations if (iii) $\eps_{\rm turb}/\eps_{\rm th}$ rises somewhat more steeply with radius. The fact that $\eps_{\rm turb}/\eps_{\rm th}$ rises with radius is well supported by cosmological hydrodynamic simulations of clusters \citep{2009ApJ...705.1129L,2010ApJ...725.1452S,vazza11}; even the last model (iii) is (within the scatter) perfectly consistent with the trends seen by such simulations. This strongly suggests that a rising $\eps_{\rm turb}/\eps_{\rm th}$ should be part of any final model of radio haloes. Given the lack of observational constraints on turbulent profiles, this unfortunately means that models will have yet another degree of freedom. 

The above models are merely meant to serve as existence proofs, showing what
modifications to the standard model are necessary to be reconciled with the
radio halo observations. There are too many uncertainties and parameter
degeneracies for any of these models to be definitive, or for one to make firm
quantitative statements about turbulent and seed CR profiles, apart from placing
certain bounds. We abstract the main parameter dependencies by exploring static,
spherically symmetric models consistent with Coma, where we repeat our
Fokker-Planck calculations. We explore three main effects: the overall
normalisation of turbulence, and spatial distribution of turbulence and CRs.

We find that the most important variable is the overall normalisation of
turbulence: surface brightness scales exponentially with the amount of
turbulence. The radio spectrum also becomes more curved with higher
turbulence. In contrast, the surface brightness scales linearly with the CR
abundance.  The spatial distributions of both turbulence and CRs also influence
radio halo profiles (for both, flatter distributions at fixed overall
normalisation imply flatter surface brightness profiles with lower overall radio
luminosity), but play a secondary role.

This exponential dependence of radio halo luminosity with turbulence then raises
the interesting question of why radio halo scaling relations (e.g.,
$L_{\nu}$--$L_{\rm X}$) are so tight. Previous statistical modelling of radio
halo populations \citep{2006MNRAS.369.1577C,2007MNRAS.378.1565C} have focused
on matching the mean relations, but not the scatter. We believe the latter
provides a particularly interesting and potentially fruitful constraint. The
acceleration timescale $\tau_{\rm D}$ and the length of time during which
acceleration takes place $\tau_{\rm cl}$ must be matched so that there are
roughly $\sim 2$ e-folds (given the factor $\sim 10$ difference in radio
luminosity between the radio bright and faint populations; \citet{brown11}),
otherwise scatter in turbulence (which we might expect given the wide variety of
merger and infall conditions) will be exponentially amplified in radio
luminosity. The timescale of the merger has some natural scatter which is not
obviously correlated with $\tau_{\rm D}$. 

If instead we adopt $\tau_{\rm cl}/\tau_{\rm decay} \sim\rmn{const}$,
i.e. the natural lifetime of turbulence is its decay lifetime, then we can show
that the acceleration time and the lifetime of turbulence are coupled, simply
because the wave-particle interaction rate (which drives particle acceleration)
and the wave-wave interaction rate (which drives dissipation) are linked. If we
are mindful that Kraichnan turbulence is only applicable below the Alfven scale
$l_{\rm A}$ (where turbulent velocities $\varv < \varv_{\rm A}$), then
remarkably the ratio $\tau_{\rm cl}/\tau_{\rm D}$ is independent of properties
of the turbulence (such as its amplitude and outer scale) and only depends on
plasma parameters. Our results suggest that TTD on thermal particles may result
in overly rapid damping of turbulence, and inefficient acceleration (equation
\ref{eqn:ratios_TTD}), but if plasma instabilities scatter the thermal particles
so that the fast modes damp via TTD on the CRs, then the required acceleration
can take place (equation \ref{eqn:ratios_CR_damp}). In agreement with other
assessments, we find that if turbulence is Burgers rather than Kraichnan, then
turbulent reacceleration is ineffective, and some other explanation for radio
haloes (e.g., magnetic reconnection, \citealt{brunetti16}) is necessary. If we
adopt the ansatz $\tau_{\rm cl}/\tau_{\rm decay} \sim\rmn{const}$, then we
find that above a threshold level of turbulence (necessary to overcome cooling),
radio properties are insensitive to both the amplitude and spatial profile of
turbulence, since only a fixed, asymptotic amount of amplification takes
place. On the other hand, they {\it are} sensitive to the properties of the seed
electron profile. This raises the exciting possibility that one could
marginalise over the highly uncertain properties of cluster turbulence to learn
about the underlying seed CR population. Our results suggest that studying not
just mean trends, but the relatively small scatter in radio-halo scaling
relations, may prove extremely fruitful in shedding light on ICM plasma
processes.

{\bf Acknowledgements.} We thank Josh Wiener for discussions on CR streaming,
Lawrence Rudnick for discussion on uncertainties in the 1.4 GHz radio data,
Fabio Zandanel for re-calculating gamma-ray limits, and Gianfranco Brunetti for
useful discussions. We thank an anonymous referee for helpful comments that
improved the paper. A.P. is grateful to the Swedish research council for
financial support. S.P.O. thanks NASA grants NNX12AG73G and NNX15AK81G for
support, as well as KITP for hospitality. This research was supported in part by
the National Science Foundation under Grant No. NSF
PHY11-25915. C.P.~acknowledges support by the European Research Council under
ERC-CoG grant CRAGSMAN-646955 and by the Klaus Tschira Foundation.

\vspace{-0.7cm}

\bibliography{paper}
\bibliographystyle{mnras}

\end{document}